\documentclass[preprint,authoryear,12pt]{elsarticle}

\usepackage{float}
\usepackage{hyperref}
\usepackage{xurl}
\usepackage{amssymb}

\usepackage{pifont}
\usepackage{xcolor}
\newcommand{\cmark}{\textcolor{green}{\ding{51}}}
\newcommand{\xmark}{\textcolor{gray}{\ding{55}}}
\newcommand{\partmark}{\textcolor{orange}{\textbf{$\sim$}}}
\usepackage{makecell}
\usepackage{amsmath}
\usepackage{multirow}
\usepackage{listings}
\journal{Computer \& Education}

\begin{document}

\begin{frontmatter}

\title{Can AI be a Teaching Partner? Evaluating ChatGPT, Gemini, and DeepSeek across Three Teaching Strategies}

%\author{Anonymous for double-blind review}

%% Author 1
\author[label1,label3]{Talita de Paula Cypriano de Souza}
\ead{talita@ifsp.edu.br}
\affiliation[label1]{
    organization={Federal Institute of São Paulo (IFSP)},
    address={Major Fernando Valle Avenue, 2013},
    city={Braganca Paulista},
    postcode={12903-000},
    state={SP},
    country={Brazil}
}

%% Author 2
\author[label2]{Shruti Mehta}
\ead{shrmeh@upenn.edu}
\affiliation[label2]{
    organization={Graduate School of Education of University of Pennsylvania (UPENN)},
    address={3700 Walnut Street Philadelphia, PA 19104},
    city={Philadelphia},
    postcode={610101},
    state={Pennsylvania},
    country={United States}
}

%% Author 3
\author[label3]{Matheus Arataque Uema}
\ead{matheuema@usp.br}
\affiliation[label3]{
    organization={Institute of Mathematics and Computer Sciences of University of São Paulo (USP)},
    address={Trabalhador São-Carlense Avenue, 400},
    city={Sao Carlos},
    postcode={13566-590},
    state={Sao Paulo},
    country={Brazil}
}

%% Author 4
\author[label1]{Luciano Bernardes de Paula}
\ead{lbernardes@ifsp.edu.br}

%% Author 5
\author[label2,label4]{Seiji Isotani\corref{cor1}}
\cortext[cor1]{Corresponding author}
\ead{sisotani@upenn.edu}
\affiliation[label4]{
    organization={Nucleus of Excellence in Social Technologies (NEES) of Federal University of Alagoas (UFAL)},
    address={Lourival Melo Mota Avenue, S/N},
    city={Maceió},
    postcode={57072-900},
    state={Alagoas},
    country={Brazil}
}

%% Abstract
\begin{abstract}
There are growing promises that Large Language Models (LLMs) can support students’ learning by providing explanations, feedback, and guidance. However, despite their rapid adoption and widespread attention, there is still limited empirical evidence regarding the pedagogical skills of LLMs. This article presents a comparative study of popular LLMs, namely, ChatGPT, DeepSeek, and Gemini, acting as teaching agents. An evaluation protocol was developed, focusing on three pedagogical strategies: Examples, Explanations and Analogies, and the Socratic Method. Six human judges conducted the evaluations in the context of teaching the C programming language to beginners. The results indicate that LLM models exhibited similar interaction patterns in the pedagogical strategies of Examples and Explanations and Analogies. In contrast, for the Socratic Method, the models showed greater sensitivity to the pedagogical strategy and the initial prompt. Overall, ChatGPT and Gemini received higher scores, whereas DeepSeek obtained lower scores across the criteria, indicating differences in pedagogical performance across models.
\end{abstract}


%% Keywords

\end{frontmatter}



\section{Introduction}\label{sec:int}
Large Language Models (LLMs) such as ChatGPT, Claude, and Gemini have rapidly expanded to everyday use, including educational contexts \citep{gptEducation2023, google-learn-2025, case-study, wang2025chatgpt}. Teachers and students have increasingly adopted these systems to enhance multiple aspects of the educational process, including adaptive learning facilitation, personalized feedback provision, and the simulation of tutoring interactions \citep{mollick2023, roleGpt2023}.

As the adoption of LLMs grows, benchmarks have been gaining traction for assessing the model's performance, strengths, and limitations \citep{arena2024, mmlu2021, livebench2024, benchped2025, evaluatingbehaviors2024}. 
However, while emerging benchmarks have begun to assess educational capabilities \citep{mrbench2025,math-tutor-2025,eli-why-2025}, a critical gap persists regarding how models adapt strategies during real-time interaction with students. Here, we define pedagogical strategies as the set of instructional decisions and actions undertaken by a teaching agent to guide the learning process \citep{multi-level}.

Educational theories highlight the role of examples \citep{vanGog2010, examples97, examples2000}, explanations, and analogies \citep{mollick2023strategies} as fundamental strategies for scaffolding student learning. In addition, the Socratic Method fosters critical thinking through guided questioning \citep{socratic2016}. Although these approaches have been widely used in human teaching and learning, it remains unclear whether LLMs can adopt them in ways that meaningfully support student learning. Clarifying this distinction is crucial because it distinguishes merely providing correct answers from actively facilitating learning, with significant implications for educational practice.

Furthermore, investigating the pedagogical potential of LLMs contributes to ongoing discussions on the ethical and social dimensions of artificial intelligence in education \citep{gptEducation2023}. This perspective underscores the need to evaluate these tools from an educational standpoint, thereby identifying their limitations and biases and validating their potential contributions to teaching and learning environments. In this regard, previous studies \citep{sbie2023, rbie2025} analyzed the quality of LLM-generated responses in the context of teaching programming to beginners. The studies initially analyzed the style and correctness of LLMs' responses to C programming language exercises. The models demonstrated a strong ability to provide correct answers; however, in several cases, prior knowledge was required to assess the response adequately. These findings also indicate that the models tended to generate overly complex explanations for novice-level questions.

Understanding whether these systems can teach, not just respond, is particularly relevant in contexts marked by educational inequalities, where AI tools may serve as complementary resources to mitigate gaps in access to qualified instruction. 
This concern becomes even more pressing in resource-limited settings, as highlighted by \cite{unplugged2023} in their definition of Artificial Intelligence in Education (AIED) Unplugged. From this perspective, AI can provide individualized support to students, especially in contexts where a teacher is unavailable \citep{google_ai_future_learning}. However, this potential also underscores the need to evaluate how LLMs can be effectively and responsibly deployed to support learning in diverse environments where traditional pedagogical resources are scarce \citep{frame-unplugged}. 

In this context, this study aims to answer the following research question: Are LLMs capable of teaching using specific pedagogical approaches? To investigate this, we designed an evaluation protocol that analyzes LLM outputs across three pedagogical strategies: Examples, Explanations and Analogies, and the Socratic method, applied to the context of teaching the C programming language. This work contributes to the field of AIED by analyzing the pedagogical behaviors of well-known LLMs. It seeks to move beyond accuracy-based benchmarks by proposing a framework to assess the educational value of LLM interactions and their potential to support teaching and learning processes in different contexts.

The remainder of this paper is organized as follows. Section 2 presents related work, contextualizing the general and educational benchmarks of LLMs while underscoring the need to investigate the pedagogical skills of LLM models. Section 3 details the methodology, with emphasis on the design of the evaluation protocol, the adopted criteria, and the selection of LLMs. Section 4 reports the evaluation results across the pedagogical strategies, and Section 5 discusses the limitations of the research. Finally, Section 6 concludes the paper by summarizing its contributions and indicating opportunities to advance the evaluation of LLMs in educational contexts, followed by an ethics statement in Section 7. Supplementary materials, including Colab notebooks, datasets, and scripts used in this study, are publicly available at \textit{[omitted for double-blind review]}, ensuring reproducibility and extensibility of our experiments.

\section{Related Work}\label{sec:rel}

\subsection{Benchmarks for LLMs and Pedagogy}
The rapid advancement of LLMs has spurred the development of numerous benchmarks. These evaluation frameworks use a combination of test datasets and performance metrics to systematically compare the capabilities and risks of AI models \citep{raji2021ai}. This process not only fosters model innovation but also helps select the most suitable ones for specific contexts. The sources of the questions and the evaluation metrics are crucial factors considered in these methodologies. 
Several prominent benchmarks exemplify the current approach. Chatbot Arena \citep{arena2024} is an open platform for live evaluation based on human preferences, where users vote for the best response between two anonymous LLMs. The Measurement of Massive Multitask Language Understanding (MMLU) \citep{mmlu2021} is a benchmark designed to evaluate the accuracy of LLMs across a wide range of tasks and knowledge domains, including elementary mathematics, US history, and computer science. LiveBench \cite{livebench2024} addresses challenges such as test-set contamination and the limitations of LLM judges or crowdsourcing.

While these works are fundamental for assessing general performance, they do not address the specificities and nuances of the educational context. This gap is a critical point highlighted by a preliminary report from \cite{aifored2024}. The report identified significant shortcomings in existing evaluations, including the need to incorporate aspects of real-world engagement by teachers and students with tools (including their behaviors and preferences) and the absence of metrics to assess pedagogical skills and curriculum-aligned content.

Addressing this domain, \cite{bechnmarkingknow} introduces the Cross-Domain Pedagogical Knowledge (CDPK), also known as the Pedagogy Benchmark, designed to evaluate whether LLMs possess pedagogical knowledge and can support student learning. The authors built this benchmark by extracting questions from an annual national exam used to assess teachers in Chile, modeling it to determine the ability of LLMs to provide correct answers. The tested models achieved accuracy rates ranging from 28\% to 87\%. An important point highlighted by the authors is that the benchmark does not evaluate real classroom applications, which they identify as a potential direction for future research. In addition, the test is limited to verifying whether LLMs can pass teacher training exams, with pedagogical knowledge distributed across different subject areas and education levels. In this sense, the work presented in our paper aims to complement the work by \cite{bechnmarkingknow} by exploring the pedagogical skills of LLMs in practice, that is, how they behave when engaged in actual learning tasks.

Similarly, \cite{evaluatingbehaviors2024} argue that traditional benchmarks fail to reflect high-stakes scenarios in educational usage. They propose a framework for model behavior evaluation with the following principles: knowledge, robustness, taxonomy, negation, and fairness. For their use case, the authors evaluated the behavior of LLMs in student-written responses to science questions. Despite being a robust framework that considers aspects not addressed by other benchmarks, the focus remains on how LLMs behave when assessing student assignments.

More recently, \cite{edubench2025} propose EduBench, a benchmark focused on evaluating LLMs in diverse educational scenarios. EduBench fills in the gaps left by previous evaluations that focused on purely knowledge-intensive tasks and did not consider pedagogical relevance. The proposed scenarios are categorized as student-centered, teacher-centered, and general applications. The data is synthetically generated, and a set of metrics is used for evaluation. Despite the well-explored contributions and scenarios, it lacks a focus on the model's pedagogical approach.

These related works highlight the relevance of developing benchmarks for LLMs, as well as the challenges and opportunities that remain for investigation, given the complexity and diversity of the educational context.

Thus, the focus should be not only on assessing correct answers but also on how those answers are provided and what pedagogy is being used. In the teaching and learning process, immediately giving the final answer does not add value to the student. Therefore, it is desirable for the LLM to be able to guide the process by understanding the student's prior knowledge and adapting its response to promote learning. This desirability for LLM to guide teaching and learning calls for a better understanding and benchmarking of pedagogical skills by LLMs. 

In this context, several recent studies have addressed the evaluation of LLMs within educational contexts. In the domain of mathematics, \cite{mrbench2025} investigated the efficacy of LLMs as AI Tutors, focusing on the pedagogical strategies required for effective educational dialogue. Similarly, \cite{math-tutor-2025} introduced a benchmark with a suite of datasets and metrics to assess holistic tutoring models in mathematical interactions. 
Regarding conversational capabilities, \cite{education-q-2025} presented a multi-agent dialogue framework to evaluate the learning potential of LLMs with a focus on formative assessment in simulated educational scenarios. Complementarily, \cite{eli-why-2025} outlined a benchmark that evaluated the pedagogical utility of explanations, specifically analyzing the models' ability to adapt answers to 'Why' questions. Finally, in the context of programming education, \cite{traver-2025} proposed an agent workflow to analyze multi-turn tutoring interactions for coding tasks.

Table \ref{tab:related} compares key features of related works with our proposal. As observed, existing studies predominantly rely on static datasets or automated simulations (using LLMs or multi-agent systems) to assess pedagogical aspects. In contrast, our proposal employs a dynamic evaluation process where human judges interact with LLMs in real time. Regarding human evaluation, while often cited in the literature as a scalability bottleneck, we argue that it is essential for ensuring methodological rigor and approximating real-world educational scenarios (ecological validity). Furthermore, our approach emphasizes multi-turn interactions, evaluating the model's ability to sustain a pedagogical dialogue beyond a single exchange.

\begin{table}[ht]
\centering
\caption{Comparison of key features of related works and the proposed experiment.}
\label{tab:related}
\resizebox{\textwidth}{!}{%
\begin{tabular}{l l c c l l}
\hline
\textbf{Reference} & \textbf{Domain} & \textbf{\makecell{Human\\Ass.}} & \textbf{\makecell{Multi-\\turn}} & \textbf{Interaction} & \textbf{Approach} \\ \hline
\cite{edubench2025} & General & \partmark & \cmark & Simulation & Benchmark \\ \hline 
\cite{mrbench2025} & Math & \cmark & \xmark & Dataset & Benchmark \\ \hline
\cite{math-tutor-2025} & Math & \xmark & \xmark & Dataset & Benchmark \\ \hline
\cite{education-q-2025} & General & \partmark & \cmark & Simulation & Framework \\ \hline
\cite{eli-why-2025} & STEM/non STEM& \partmark & \xmark & Dataset & Benchmark \\ \hline
\cite{traver-2025} & Coding & \partmark & \cmark & Simulation & Workflow \\ \hline
\textbf{Our proposal} & \textbf{Coding} & \textbf{\cmark} & \textbf{\cmark} & \textbf{Real time} & \textbf{Protocol} \\ \hline
\end{tabular}%
}
\smallskip
\footnotesize
\raggedright
\textit{Legend: \cmark~= Presence; \partmark~= Partial; \xmark~= Absence.}
\end{table}

\subsection{Pedagogical Skills and role of LLMs}

Several studies have examined whether popular LLMs such as ChatGPT, DeepSeek, and Gemini possess sufficient knowledge to function as guides or teachers in educational contexts \citep{bechnmarkingknow, mollick2023strategies}. However, the extent to which these models can emulate the pedagogical skills of human teachers remains underexplored.

Pedagogical skill refers to the ability to translate subject knowledge into meaningful learning experiences through explanation, questioning, adaptation, and reflection \citep{multi-level}. These skills involve designing and facilitating instruction that scaffolds and supports learners, enabling them to think and learn independently.

Teaching is a complex process that requires not only content knowledge and pedagogical knowledge, but also the instructional skills necessary to implement them effectively \citep{jeschke2021}. While content knowledge involves understanding the subject matter, pedagogical knowledge involves knowing how to teach it. However, instructional or pedagogical skills represent the application of this knowledge. Pedagogical skills may be interpreted as ``doing'' that follows from ``knowing'' — such as planning lessons, managing classrooms, providing feedback, and adapting instruction in real time.

According to \cite{shulman1987knowledge}, Pedagogical Content Knowledge integrates content and pedagogy into an understanding of how to organize, represent, and adapt instruction to meet learners’ needs. One way this understanding is implemented in teaching activities is through \textit{instruction}. Instruction, as a key component of pedagogical reasoning and action, draws upon a variety of approaches, including lectures, demonstrations, cooperative learning, Socratic dialogue, project-based learning, and experiences beyond the classroom.

A comprehensive evaluation of teaching capabilities must go beyond content and pedagogical knowledge to assess the practical application of teaching methods—the pedagogical skills themselves \citep{bechnmarkingknow, jeschke2021}.

Therefore, this paper aims to address this gap by evaluating how LLMs support students in completing programming exercises. Specifically, we assess the pedagogical skills of LLMs, such as the ability to provide scaffolding, generate explanations, and employ Socratic questioning, by benchmarking models like ChatGPT, DeepSeek, and Gemini.

\section{Methodology}\label{sec:met}

Considering the widespread use of LLMs in educational contexts, the evaluation protocol was developed based on the use of prompts to guide the models' behavior in subsequent interactions. This approach is easily replicable in real-world scenarios, where teachers could instruct students to use prompts before posing questions.

For this study, three pedagogical approaches were selected for evaluation across the chosen LLMs. The protocol defined specific criteria to reduce subjectivity in the judgment process.

The tests were conducted in the context of teaching the C programming language, which is commonly used in Computer Science and Engineering courses as an introductory programming language. It provides essential foundations for control structures, memory manipulation, and computational logic. To match the expected initial level of learning and facilitate reproducibility, exercises were selected from basic topics, including operators, decisions, loops, arrays, and functions. All exercises were taken from \cite{backes2013linguagem} and are listed in  \ref{app:exercises}.

\begin{figure}[H]
\centering
\includegraphics[width=0.9\textwidth]{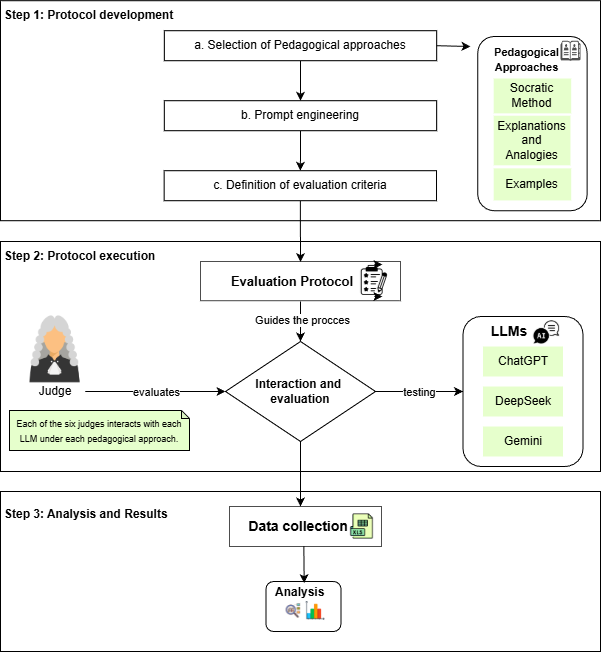}
\caption{Experimental Workflow of LLM Interaction and Evaluation}\label{fig:diagram}
\end{figure}

Six judges performed the evaluations between April and July 2025, interacting with each LLM for each pedagogical approach. Each evaluation involved a judge using a prompt corresponding to the pedagogical strategy and a C language exercise. Figure \ref{fig:diagram} illustrates the experimental workflow adopted in this study.

\subsection{Pedagogical Approaches}

\cite{mollick2023strategies} present five strategies that teachers can use with LLMs: Examples, Explanations and Analogies, Low-Stakes Tests, Assessment of Knowledge Gaps, and Distributed Practice. Each strategy includes a defined prompt and a set of criteria to evaluate whether the LLM’s output is suitable for student use. For the development of the protocol presented in this article, we selected two of these strategies, Examples and Explanations and Analogies, because they allow for direct interaction between students and the LLM. Additionally, the Socratic Method was included due to its widespread use by teachers in classroom settings \citep{socratic2016}.

The Examples strategy aims to improve the student's understanding of a concept \citep{examples97, examples2000}, fostering critical thinking, analysis, evaluation, and the ability to generalize concepts and principles to new situations \citep{mollick2023strategies}. 
In this study, when a student presents a question or a description of a programming exercise, the LLM should not provide an immediate full solution. Instead, it offers a code example that illustrates the relevant function. For instance, given the exercise: ``\textit{Write a program that reads four float values, calculates, and displays the arithmetic mean of these values}'' an appropriate example could be a snippet that reads two values, performs an operation, and displays the result:

\begin{lstlisting}
float n1, n2, result;
scanf("%f", &n1);
scanf("%f", &n2);
result = n1 + n2;
printf("Result:%.2f\n", result);
\end{lstlisting}

Based on this example, students can remember how to declare variables, assign a value, perform an arithmetic operation, and display the result. Thus, students from the given example could develop their own solution, while the LLM provides additional examples as new questions arise.

The Explanations and Analogies strategy offers a structured knowledge base to help students organize information and understand concepts. Analogies provide concrete representations that link prior knowledge with new content, and when chosen appropriately, they promote concept learning by encouraging students to connect familiar experiences with new contexts and problems \citep[][chap.~5]{analogies2006}. 
In a C programming context, if a student presents an exercise such as ``\textit{Write a program that reads an array with eight positions, reads two values X and Y corresponding to array positions, and shows the sum of values at positions X and Y}'', the LLM could explain arrays using an analogy such as ``the drawers of a dresser'', where each drawer represents an array position.

The Socratic Method, inspired by Socrates (469–399 BC), promotes deep understanding through structured dialogues. Teachers serve as facilitators, guiding reasoning by eliciting hypotheses and using counterexamples to clarify concepts \citep{socratic2016}. This method consists of five interdependent stages: (i) posing an initial question (wonder), (ii) formulating a provisional hypothesis, (iii) refuting and critically examining the hypothesis (elenchus), (iv) accepting or rejecting the hypothesis based on counterexamples, and (v) applying the findings in practice. 
In the programming context, previous research highlights that the Socratic Method relies on a sequence of guided questions that emphasize key aspects of the target code, which may explain its superior performance compared to freer approaches \citep{socraticProgramming}.
In our scenario, the LLM should guide the student with questions, advancing as the student’s responses prove correct. For example, given the exercise: ``\textit{Write a program that reads two integer numbers and shows which is larger}'', the LLM might initiate dialogue by asking about input functions and variables, progressing as the student formulates hypotheses. 

Ideally, teachers adopt the most suitable pedagogical approach for each type of question, concept, and student profile. However, in this experiment, each approach was tested in isolation within each LLM to ensure a controlled comparison.

\subsection{Prompting}

A strategy for prompting is essential when evaluating the behavior of LLMs. You can use different approaches, such as one-shot, few-shot, thought generation, and other techniques \citep{prompting}. Using different strategies helps ensure better results and responses that align more closely with the desired behavior. 

To evaluate LLMs in this study, we chose the ``role prompting'' strategy. In this approach, a single initial prompt is provided that includes all the rules and expected interaction formats. In addition, LLM behavior and personality are defined, e.g. ``acts as an expert in the field'' \citep{prompting}.
This strategy was adopted to simulate how a programming student might use the tool in real-life scenarios. For instance, a teacher could provide an initial prompt while the student guides the subsequent interaction, or in resource-limited settings \citep{unplugged2023}, the prompt could be offered to the student for independent study.

In this way, to obtain the expected behavior from each LLM, prompt engineering was performed as presented by \cite{mollick2023}. 
The authors propose seven ways to use LLMs in the classroom, by ``role prompting'' strategy. Among them is the role of an ``AI tutor'', which is explored in this article. Furthermore, for prompt elaboration, they indicate organizing them into a role and a goal, step-by-step instructions, pedagogy, constraints, and personalization.

Therefore, to create each prompt, every element proposed by \cite{mollick2023} was included. In this case, the LLM was configured to behave as an optimistic and encouraging teacher for students. The goal was to help students at the beginner level solve a list of programming exercises. The step-by-step process and the pedagogical approach were specific to each one selected (Examples, Explanations and Analogies, and Socratic Method). The rules were not to provide ready answers or solutions, but to understand and adapt responses according to the student's levels of understanding.

Preliminary tests were conducted to ensure that the responses of the selected LLMs (ChatGPT, DeepSeek, and Gemini) were consistent with what was requested for each pedagogical approach. 

To illustrate the experimental setup, the specific prompt used for the Examples approach is presented below. For completeness, the full prompts for all pedagogical approaches (including Explanations and Analogies and the Socratic Method) are detailed in \ref{app:prompts}.

\begin{quote}
“You are an upbeat, encouraging teacher who helps students understand concepts and solve a list of beginner-level C programming exercises. To help students solve the exercises, first help them understand the related topics by providing \textbf{examples}. You should guide students in an open-ended way. Do not provide immediate answers or solutions. Instead, help students generate their own answers by using \textbf{examples}. When a student demonstrates that they know the concept, you can conclude the conversation. Rules: Do not assume the students can accurately assess their own understanding. Your job is to assess what the student understands and adapt your \textbf{examples} to their level of understanding."
\end{quote}

\subsection{Protocol of evaluation}
To perform a comprehensive evaluation of the selected models and approaches, six judges with varying profiles but advanced knowledge of programming languages were chosen. This diverse group allowed them to critically assess the models. The differences in the judges' backgrounds also helped make the evaluations more robust and provided complementary perspectives for the analysis.

To reduce the subjectivity of the evaluation and ensure more methodological rigor, an evaluation protocol was created to standardize the judging criteria.

\begin{figure}[H]
\centering
\includegraphics[width=0.9\textwidth]{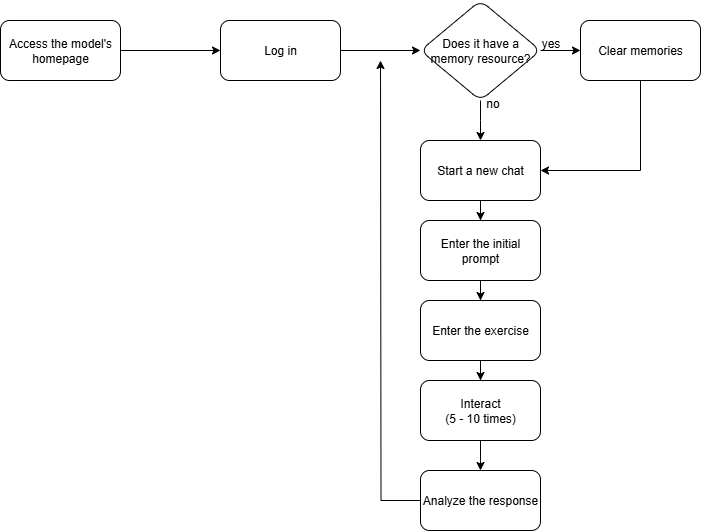}
\caption{Process of interaction with LLM adopted by each judge}\label{figProtocol}
\end{figure}

Each judge performed the interaction process with all selected LLMs, as presented in Figure \ref{figProtocol}.

First, the judge accessed the LLM homepage and logged in. Then, if the model had a memory feature like ChatGPT, the judge cleared its memory and followed the steps: start a new chat, input the initial prompt, input the exercise, and interact with the model. After finishing the interaction, the judge graded all the defined aspects of the evaluation protocol.

In the developed protocol, interaction rules were defined to keep the evaluations more consistent and less subjective. The first definition was the number of interactions that should be performed for each exercise, with each model, from 5 to 10 interactions. We established this limit to ensure that the interaction was sufficiently extensive for evaluation without overloading the model or compromising the analysis of its behavior.

Additionally, the judge interacted as if he or she were a beginner programming student, responding to each interaction in one of the following ways:

\begin{itemize}
\item Stating that he or she doesn't know how to answer what the LLM is asking;
    
 \textbf{Example:} The LLM asked the judge a question, and he or she responded with ``I don't know'' or ``I don't understand''.
    
\item Providing a wrong answer;
    
 \textbf{Example:} The LLM asked the judge a question, and he or she deliberately responded with a wrong answer.
    
\item Requesting a new explanation;
    
 \textbf{Example:} The LLM explained a concept and asked the judge a question, and he or she responded with ``explain better'' or ``explain again''.
    
\item Providing correct answers, as long as he or she uses information that the LLM itself provided in a previous explanation.
    
 \textbf{Example:} If the LLM had previously explained the concept of the successor of a number and the predecessor, and then asked what the corresponding values would be for 10, the judge should correctly answer ``9 and 11''.
\end{itemize}

\subsection{Evaluated aspects}

After defining the prompts to be used and the evaluation protocol was established, the evaluation aspects were determined. To reduce the subjectivity of the evaluation and obtain more details about the characteristics of each LLM, the aspects proposed by \cite{mollick2023strategies} were adapted to guide each judge's assessment.

To this end, after interacting with the model regarding a specific exercise in each approach (see \ref{app:questions} for the full list of guiding questions), the judge filled out a form assigning grades according to a scale of 0 to 5, where 0 is totally unsatisfactory and 5 is totally satisfactory for specific questions in each approach.

For the Examples approach, the aspects were:
\begin{itemize}
   \item \textbf{Relevance}: Whether the provided examples are directly related to the questions posed;
   
   \item \textbf{Correctness}: Whether the provided examples allow students to execute the code successfully and obtain the expected solution; 
   
   \item \textbf{Details level}: Whether the provided examples contain sufficient detail to allow independent reproduction and understanding; 
   
    \item \textbf{Variety}: Whether the provided examples cover different contexts and perspectives, rather than being restricted to a single scenario; 
    
   \item \textbf{Abstract–concrete connections}: The extent to which the provided examples establish clear links between abstract concepts and real-life applications.
   
\end{itemize}

For the Explanations and Analogies approach, the evaluating aspects were as follows:
\begin{itemize}
    \item \textbf{Explanation}: Whether a clear explanation was provided to support the understanding of the topic addressed in the exercise; 
    
    \item \textbf{Analogy}: Whether an analogy was included to facilitate comprehension by relating the topic to a familiar or comparable situation; 
    
    \item \textbf{Clarity, consistency and ease}: Whether the explanations and analogies were expressed in a clear, consistent, and accessible manner; 
    
   \item \textbf{Critical parts focus}: Whether the explanations and analogies addressed the most critical aspects of the concept, highlighting essential elements; 
    
    \item \textbf{Correctness}: Whether the explanations and analogies were accurate and aligned with the intended solution of the exercise;
    
   \item \textbf{Level adaptation}: Whether the explanations and analogies were adapted to the beginner's level of knowledge and expertise of the students;
   
    \item \textbf{Usefulness}: Whether the explanations and analogies were helpful in facilitating understanding and problem-solving;
    
   \item \textbf{Previous knowledge connection}: Whether the explanations and analogies established links with students’ prior knowledge, promoting integration with previously learned concepts.
   
\end{itemize}

For the Socratic Method approach, the aspects were as follows:
\begin{itemize}
    \item \textbf{Initial question}: Whether there was an initial question about the main topic of the exercise posed to stimulate curiosity and inquiry;
    
   \item \textbf{Counterexamples}: Whether the models presented counterexamples after an attempted answer by the student, validating or refuting it;
    
   \item \textbf{Only questions}: Whether the models rely exclusively on carefully constructed questions, rather than on direct instruction, to guide reasoning and reflection;
    
   \item \textbf{Well-formulated questions}: Whether the questions were carefully designed to encourage investigation and exploration of the topic;
   
   \item \textbf{Critical thinking promotion}: Whether the questions were able to stimulate critical thinking and deeper reasoning. 
   
\end{itemize}

In addition to these criteria for each aspect, at the end of the evaluation, the judges were required to analyze and complete their ``perceptions'', which could be ``satisfactory'', ``partially satisfactory'' or ``unsatisfactory''. 
For each approach, the judges were asked to assess whether the LLM remained within the scope of the prompt and refrained from providing an immediate solution, defined as the complete code for the exercise. This guideline was established because the expected behavior was not to deliver the correct answer directly, but rather to guide the student toward reaching the solution independently. The possible responses for this aspect were ``yes'', ``partially'' or ``no''. Finally, each judge could include a general comment about the interaction performed.

\subsection{Models}
The LLMs selected for testing were Gemini 2.5 flash, ChatGPT 4.1-mini, and DeepSeek V3. All free versions were tested on each model’s own web interface. These models were selected because they were available to students and accessible to them at the time of test execution. 

After the judges completed their evaluations, the numeric data were statistically analyzed to verify the differences among the LLMs. To assess whether the data followed a normal distribution, the Shapiro–Wilk test was applied, as it tests in small to medium-sized samples \citep{shapiro}. The results indicated a violation of the normality assumption, which justified the choice of nonparametric methods, since they do not rely on distributional assumptions. To compare the models according to the evaluation criteria, the Kruskal-Wallis test was used, as it is the non-parametric alternative to one-way ANOVA for more than two groups. In cases where statistically significant differences were detected, Dunn’s test with Bonferroni correction was used to perform pairwise comparisons while controlling for the inflation of Type I error due to multiple tests \citep{nonparametric}. For categorical variables, namely, judges' perception and immediate solution, the Chi-Square test was adopted, as it allows one to test independence and detect whether the distribution of responses differed among the models. Subsequently, to identify specific differences, the pair-to-pair comparison with the Bonferroni correction was used \citep{sharpe2015chi}. 

\section{Results and Discussion}\label{sec:res}

Based on the protocol described in Section \ref{sec:met}, each of the six judges interacted with 25 exercises for each LLM and for each pedagogical approach, following the defined evaluation criteria. Consequently, each model within each strategy received 150 evaluations. At the end of the process, the dataset comprised 1,350 evaluations, resulting from approximately 8,100 interactions (about six per evaluation) and around 270 hours of total interaction time (approximately 12 minutes per evaluation). The complete dataset and materials to enable reproducibility of the experiments are publicly available at \textit{[omitted for double-blind review]}. This section presents the results of these evaluations. 
The complete tables from the statistical analysis are available in
\ref{app:statistical}. 

\subsection{Examples Approach}
The results obtained from the evaluations of the Examples approach for each LLM are presented in Table \ref{tab:examples}. We observed that the three models presented similar average scores: ChatGPT (4.180), DeepSeek (4.035), and Gemini (4.159). This indicates that the judges evaluated the models in a relatively similar manner.
The mean and standard deviation for each evaluation criterion suggest that, for the ``Correctness'' criterion, the models obtained very similar scores as well: ChatGPT (4.927), DeepSeek (4.867) and Gemini (4.733). 
A similar pattern was observed for the criterion ``Level of detail'', with close mean scores among the models: ChatGPT (4.493), DeepSeek (4.427), and Gemini (4.220). 

A key highlight is that, for the criterion ``Variety'', all models obtained their lowest scores: ChatGPT (3.273), DeepSeek (3.093), and Gemini (3.327). This result suggests that the examples provided by the models lacked contextual variation.
For the ``Relevance'' criterion, the mean scores for the models were: ChatGPT at 4.607, DeepSeek at 4.280, and Gemini at 4.360. ChatGPT stood out, presenting the highest average, while Gemini and DeepSeek obtained quite similar scores.
In the ``Abstract-concrete connections'' criterion, ChatGPT obtained a mean score of 3.600, DeepSeek 3.507, and Gemini 4.153. Consequently, Gemini presented the highest mean score for this criterion. This suggests that the judges perceived that Gemini's examples enabled a stronger connection between concrete and abstract concepts.

\begin{table}[!htbp]
\centering
\caption{Descriptive statistics of the criteria evaluated for ChatGPT, DeepSeek, and Gemini in the Examples Approach.}
\label{tab:examples}
\renewcommand{\arraystretch}{1.2}
\setlength{\tabcolsep}{3pt}
\begin{tabular}{lccccccccc}
\multirow{2}{*}{\textbf{Criteria}} & 
\multicolumn{3}{c}{\textbf{ChatGPT}} & 
\multicolumn{3}{c}{\textbf{DeepSeek}} & 
\multicolumn{3}{c}{\textbf{Gemini}} \\
& \textbf{M} & \textbf{Md} & \textbf{SD} 
& \textbf{M} & \textbf{Md} & \textbf{SD} 
& \textbf{M} & \textbf{Md} & \textbf{SD} \\ \hline
Final Average & 4.180 & 4.400 & 0.710 & 4.035 & 4.200 & 0.928 & 4.159 & 4.400 & 0.784 \\
Relevance & 4.607 & 5.000 & 0.919 & 4.280 & 5.000 & 1.243 & 4.360 & 5.000 & 0.822 \\
Correctness & 4.927 & 5.000 & 0.286 & 4.867 & 5.000 & 0.564 & 4.733 & 5.000 & 0.808 \\
Level of detail & 4.493 & 5.000 & 0.849 & 4.427 & 5.000 & 0.951 & 4.220 & 5.000 & 1.110 \\
Variety & 3.273 & 3.000 & 1.395 & 3.093 & 3.000 & 1.688 & 3.327 & 4.000 & 1.657 \\
Abstract-concrete \\ connections & 3.600 & 4.500 & 1.932 & 3.507 & 4.000 & 1.748 & 4.153 & 5.000 & 1.441 \\
\hline
\end{tabular}
\smallskip
\textit{M = Mean; Md = Median; SD = Standard Deviation.}
\end{table}

\begin{figure}[H]
\centering
\includegraphics[width=1\textwidth]{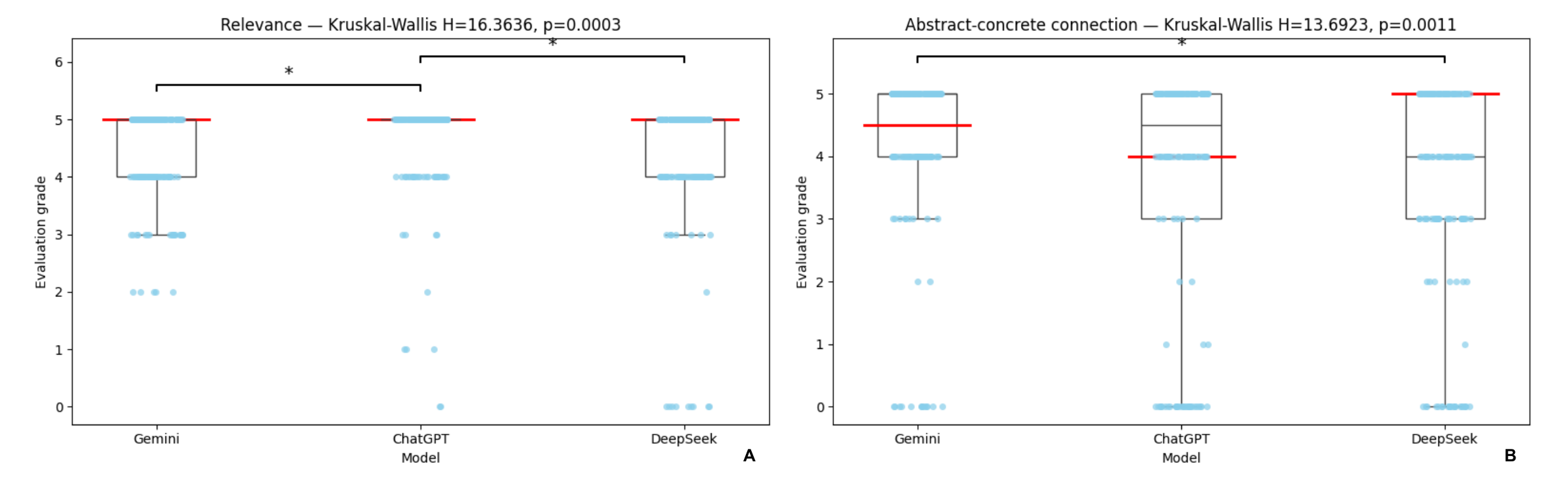}
\caption{Distribution of evaluation grades for the Examples approach based on the criteria: (A) ``Relevance'' and (B) ``Abstract-concrete connections''. Blue dots represent individual observations; boxes indicate the data distribution (quartiles), and the red line marks the median of each group — the value dividing the dataset into two equal halves and serving as a robust measure of central tendency. Black lines with asterisks denote statistically significant differences between models, identified by Dunn’s post hoc test ($p < 0.05$) following the Kruskal–Wallis test.}\label{fig:examples}
\end{figure}

Figure \ref{fig:examples} presents the comparative distribution of evaluation grades for each model, and Table \ref{tab:kruskal-examples} reports the corresponding statistical results for the evaluation criteria.
The Kruskal-Wallis test did not reveal significant differences among the three models (H = 0.8268, df = 2, p = 0.6614) for the average evaluation of the criteria. Additionally, we did not find differences among the three models for ``Correctness'' (H = 5.7260, df = 2, p = 0.0571) and ``Variety'' (H = 1.7964, df = 2, p = 0.4073) criteria.
The Kruskal-Wallis test found statistical differences between the models for ``Level of detail'' (H = 6.1636, df = 2, p = 0.0459). However, Dunn’s post hoc test did not indicate significant pairwise differences (p = 0.4073). Detailed data is shown in Table \ref{tab:dun-examples-detail} in \ref{app:statistical}.

For ``Relevance'' criterion, the Kruskal-Wallis test identified a statistically significant difference (H = 16.3636; p = 0.0003). Dunn’s test, with Bonferroni correction, indicated that DeepSeek and ChatGPT were statistically different ($p_{bonferroni}$ = 0.0051), as were ChatGPT and Gemini ($p_{bonferroni}$ = 0.0005). However, the difference between DeepSeek and Gemini was not significant, as their p-values exceeded the significance threshold (p = 1.000). These values can be found in Table \ref{tab:dun-examples-relevance} in \ref{app:statistical}. 

For the criterion of ``Abstract-concrete connections'', the Kruskal-Wallis test also revealed a statistically significant difference (H = 13.6923; p = 0.0011). Dunn's test showed that Gemini differed significantly from DeepSeek ($p_{bonferroni}$ = 0.0007), while the difference between Gemini and ChatGPT ($p_{bonferroni}$ = 0.0943), and ChatGPT and DeepSeek ($p_{bonferroni}$ = 0.3771) were not significant. The complete results are presented in Table \ref{tab:dun-examples-connections} in \ref{app:statistical}.

\begin{table}[htbp]
\centering
\caption{Kruskal–Wallis Test Results for Models in Examples Approach (df = 2; n = 150)}
\label{tab:kruskal-examples}
\begin{tabular}{lrr}
\hline
\textbf{Criterion} & \textbf{H statistic} & \textbf{p-value} \\
\hline
Final Average & 0.8268 & 0.6614 \\
Relevance$^*$ & 16.3636 & 0.0003 \\
Correctness & 5.7260 & 0.0571 \\
Level of detail & 6.1636 & 0.0459 \\
Variety & 1.7964 & 0.4073 \\
Abstract-concrete connections$^*$ & 13.6923 & 0.0011 \\
\hline
\end{tabular}

\smallskip
\textit{$^*$ Variables that exhibited a statistically significant difference.}
\end{table}

\begin{figure}[H]
\centering
\includegraphics[width=0.70\textwidth]{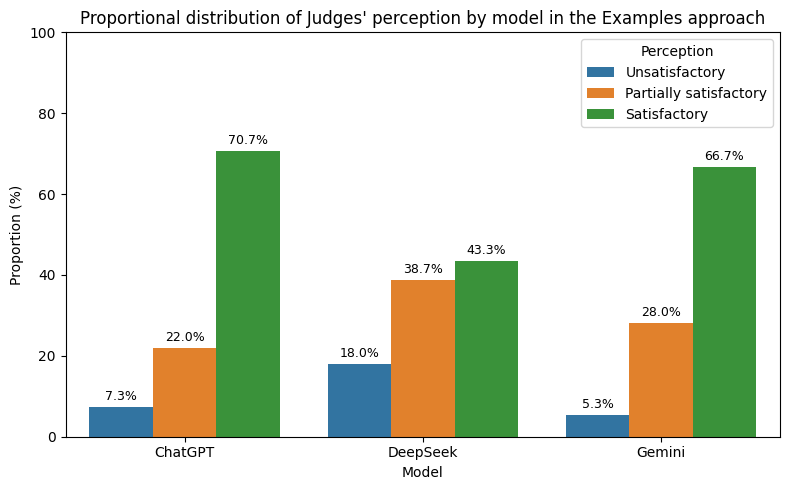}
\caption{``Judges' perception'' of Examples approach}\label{fig:examples-decision}
\end{figure}

Analyzing the ``Judges’ perception'' criterion (Figure \ref{fig:examples-decision}), we find that DeepSeek was evaluated with less satisfactory ratings compared to the other models (``satisfactory'' = 43.33\%; ``partially satisfactory'' = 38.67\%; ``unsatisfactory'' = 18\%). 
The Chi-square test revealed a statistically significant difference ($\chi^2 = 31.6979$ and $p < 0.001$). 
To identify pairwise differences among the LLMs, the Chi-square test with Bonferroni correction was applied. As presented in Table \ref{tab:dun-examples-perception} in \ref{app:statistical}, significant differences were found between ChatGPT and DeepSeek ($p < 0.001$; $p_{Bonferroni} < 0.001$) and between Gemini and DeepSeek ($p < 0.001$; $p_{Bonferroni} = 0.001$). In contrast, no significant differences were observed between ChatGPT and Gemini (p = 0.4214; $p_{Bonferroni} = 1.0000$).

\begin{figure}[H]
\centering
\includegraphics[width=0.70\textwidth]{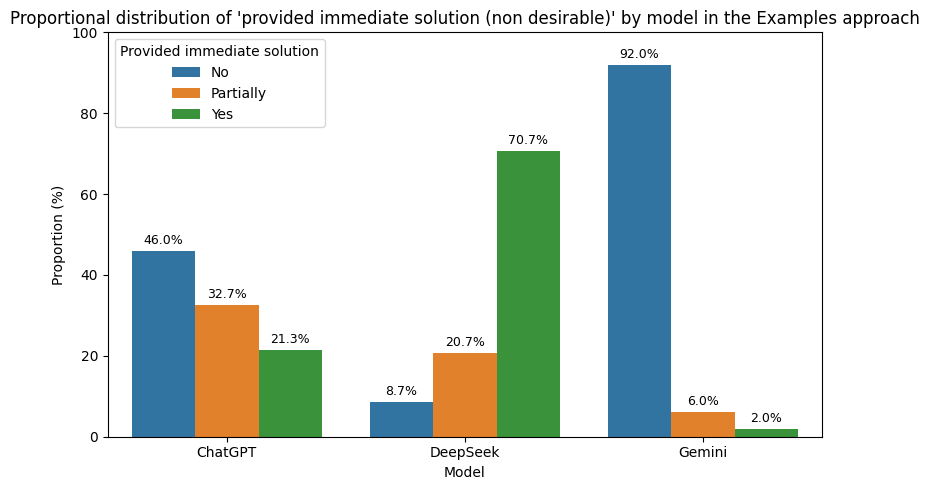}
\caption{``Provided immediate solution (non desirable)'' criterion of Examples approach}\label{fig:examples-answer}
\end{figure}

The results on the undesirable behavior of providing immediate solutions by the models are shown in Figure \ref{fig:examples-answer}. We observed that Gemini produced fewer immediate solutions and was therefore evaluated as more satisfactory for this criterion. 
Specifically, in 92\% of the evaluations it did not provide an immediate solution, in 6\% it ``partially'' did so, and only in 2\% it provided one. 
In contrast, DeepSeek was the model that most frequently produced an immediate solution, with 70.7\% of cases marked as ``yes'' 20.7\% as ``partially'' and only 8.7\% where it did not provide an immediate solution. 
ChatGPT presented intermediate results, producing an immediate solution in 21.3\% of cases (``yes’’), ``partially'' in 32.7\%, and not at all in 46.0\%.
Finally, the Chi-square test revealed a statistically significant difference ($\chi^2$ = 254.0169; $p < 0.001$). Post hoc comparisons with Bonferroni correction indicated that all pairwise comparisons (ChatGPT and DeepSeek, ChatGPT and Gemini, Gemini and DeepSeek) between models were statistically significant ($p < 0.001$). These values are presented in Table \ref{tab:dun-examples-solution} in \ref{app:statistical}.

\begin{figure}[H]
\centering
\includegraphics[width=1.0\textwidth]{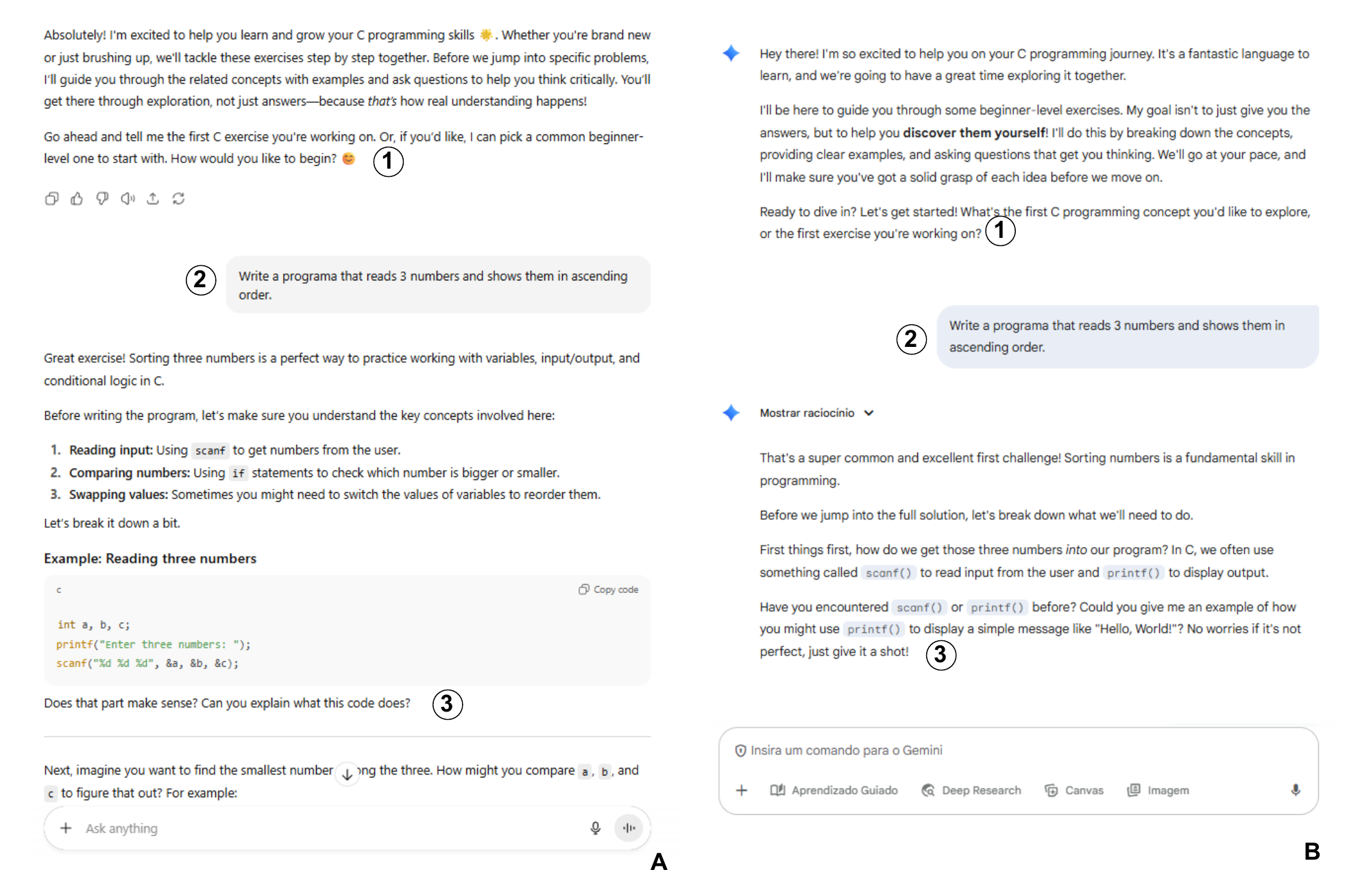}
\caption{Interaction in Examples approach: (A) ChatGPT 4.1-mini; (B) Gemini 2.5 flash.}\label{fig:chatgpt-gemini-examples}
\end{figure}

In addition to data analysis, the observations of the judges also highlighted some aspects of each model. Overall, the judges had a similar user experience with the ChatGPT and Gemini models. Conversations were more effectively guided by models that provided shorter and more direct responses. Figure \ref{fig:chatgpt-gemini-examples} presents the responses of ChatGPT (A) and Gemini (B) after the initial prompt and the description of the exercise. 
As can be observed, after the initial prompt, both models provided a direct and enthusiastic response about how they would behave (step 1). A difference is noted; after the description of the exercise (step 2), ChatGPT presented key concepts and an example of the use of reading numbers (step 3), and Gemini mentioned the commands of reading and of displaying numbers, but, in addition to giving an example, asked the user for one. 

The DeepSeek model, on the other hand, behaved differently. In many cases, it demonstrated and simulated what its behavior would be like. This interaction style can be seen in Figure \ref{fig:deepseek-simulation} in the \ref{app:interactions}, as this was the response of the model immediately after the prompt was provided. After outlining what its behavior would be, it presented an ``example flow''. In Figure \ref{fig:deepseek-simulation2} in the  \ref{app:interactions}, it is clear that the model simulated a long example dialog right after the initial prompt. This pattern was also repeated in the evaluations of the other approaches.

\subsection{Explanations and Analogies Approach}

The results obtained from the evaluations of the Explanations and Analogies approach for each LLM are presented in Table \ref{tab:explanations}. Exactly as in the previous analysis, the final average score for each evaluation was calculated.
Gemini (4.347), in turn, presented scores very close to DeepSeek (4.364). Therefore, it is possible to state that ChatGPT (4.591) presented more satisfactory scores. 
In ``Clarity, consistency, and ease'' criterion, ChatGPT (4.607) had a higher mean than DeepSeek (4.340) and Gemini (4.387). Thus, ChatGPT was evaluated by judges as providing better explanations and clearer, more consistent, and easier analogies, primarily distinguishing it from DeepSeek. 
In ``Critical parts focus'' aspect, ChatGPT (4.880) stood out and had a higher mean than DeepSeek (4.273) and Gemini (4.540). Despite that, DeepSeek and Gemini presented similar scores. 
The criterion of ``Correctness'' demonstrates that the judges perceived ChatGPT (4.893), DeepSeek (4.840) and Gemini (4.693) with the correct explanations and analogies and were evaluated similarly. 

In the ``Level adaptation'' criterion ChatGPT had the highest mean (4.660), followed by Gemini (4.540) and DeepSeek (4.360). Therefore, the data suggest that DeepSeek received lower scores for this criterion, meaning that, in the opinion of the judges, its explanations and analogies were less adapted to the student's level compared to those of ChatGPT and Gemini. 
In the ``Usefulness'' aspect, ChatGPT (4.707) was the most notable with higher scores, while DeepSeek (4.233) and Gemini (4.100) had lower and similar values. This indicates that the evaluators perceived the explanations and analogies provided by ChatGPT to be more useful compared to the other models.
The criterion of ``Connection to previous knowledge'' presented similar grades: ChatGPT (3.800), DeepSeek (4.140), and Gemini (3.820). In this case, we observe that the judges evaluated the models in a similar way in this aspect. Furthermore, all models presented their lowest means in this aspect; in other words, according to the judges, the LLMs did not provide explanations that adequately considered the student's previous knowledge.

\begin{table}[H]
\centering
\caption{Descriptive statistics of the criteria evaluated for ChatGPT, DeepSeek, and Gemini in Explanations and Analogies Approach}
\label{tab:explanations}
\renewcommand{\arraystretch}{1.2}
\setlength{\tabcolsep}{3pt}
\begin{tabular}{lccccccccc}
\hline
\multirow{2}{*}{\textbf{Criteria}} & 
\multicolumn{3}{c}{\textbf{ChatGPT}} & 
\multicolumn{3}{c}{\textbf{DeepSeek}} & 
\multicolumn{3}{c}{\textbf{Gemini}} \\ 
& \textbf{M} & \textbf{Md} & \textbf{SD}
& \textbf{M} & \textbf{Md} & \textbf{SD} 
& \textbf{M} & \textbf{Md} & \textbf{SD}  \\ \hline
Final average & 4.591  & 4.667 & 0.413 & 4.364 & 4.667 & 0.645 & 4.347 & 4.667 & 1.016 \\
Clarity, consistency, \\ and ease & 4.607 & 5.000 & 0.612 & 4.340 & 5.000 & 0.866 & 4.387 & 5.000 & 1.157  \\
Critical parts focus & 4.880 & 5.000 & 0.383 & 4.273 & 5.000 & 0.969 & 4.540 & 5.000 & 1.156  \\
Correctness & 4.893 & 5.000 & 0.369 & 4.840 & 5.000 & 0.465 & 4.693 & 5.000 & 1.042  \\
Level adaptation & 4.660 & 5.000 & 0.633 & 4.360 & 5.000 & 0.900 & 4.540 & 5.000 & 1.103 \\
Usefulness & 4.707 & 5.000 & 0.619 & 4.233 & 5.000 & 1.195 & 4.100 & 5.000 & 1.230 \\
Previous knowledge \\connection & 3.800 & 5.000 & 1.828 & 4.140 & 5.000 & 1.226 & 3.820 & 5.000 & 1.907\\
\hline
\end{tabular}
\smallskip
\textit{M = Mean; Md = Median; SD = Standard Deviation.}
\end{table}

\begin{figure}[H]
\centering
\includegraphics[width=1\textwidth]{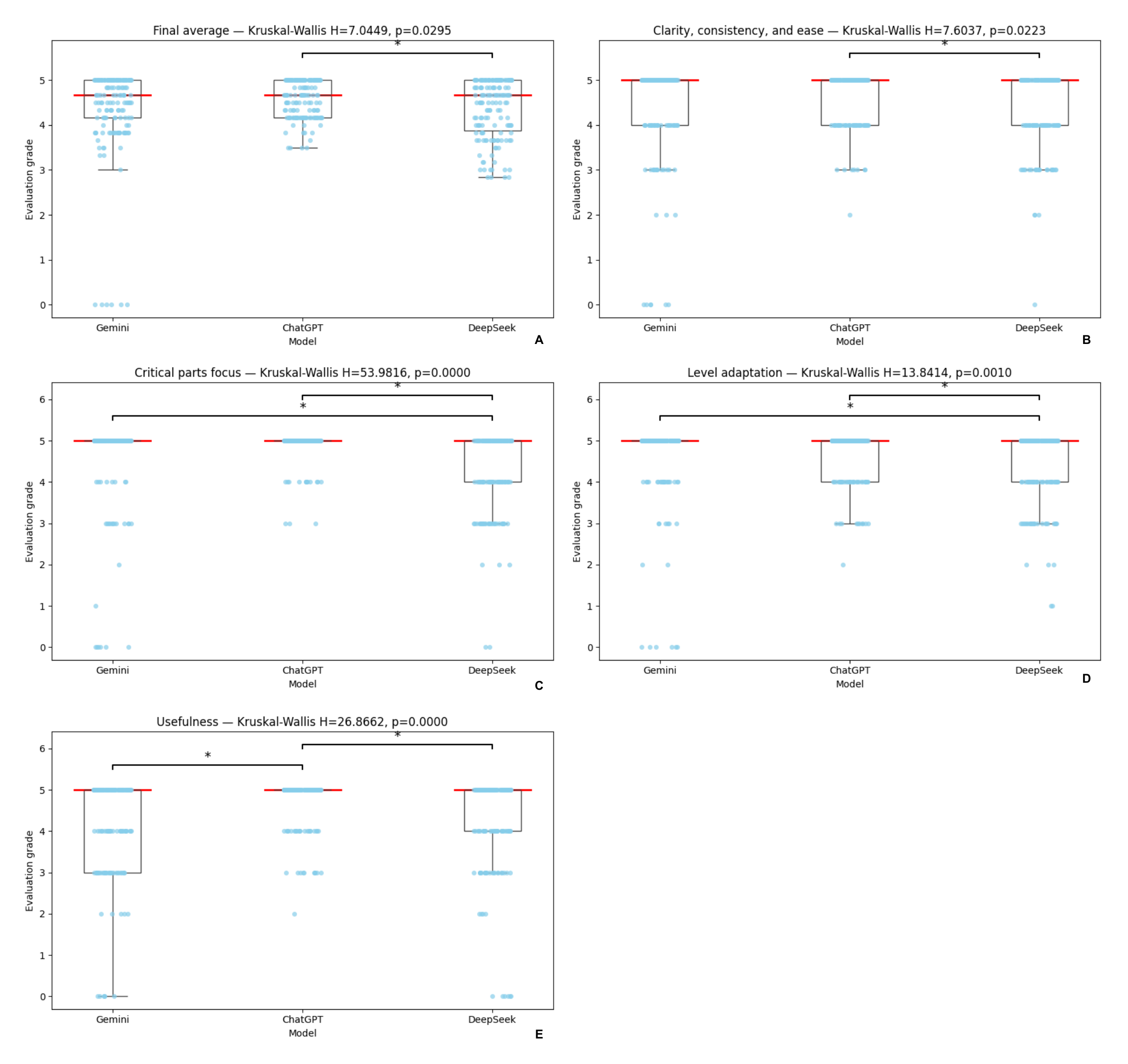}
\caption{Distribution of evaluation grades for Explanations and Analogies approach based on the criteria: (A)``Final average''; (B) ``Clarity, consistency, and ease''; (C) ``Critical parts focus''; (D) ``Level adaptation''; (E) ``Usefulness''. Blue dots represent individual observations; boxes indicate the data distribution (quartiles), and the red line marks the median of each group — the value dividing the dataset into two equal halves and serving as a robust measure of central tendency. Black lines with asterisks denote statistically significant differences between models, identified by Dunn’s post hoc test ($p < 0.05$) following the Kruskal–Wallis test.} \label{fig:explanations}
\end{figure}

Table \ref{tab:kruskal-explanations} summarizes the results of the statistical analysis, while Figure \ref{fig:explanations} illustrates the distributional differences across models.
In the ``Correctness'' criterion, Kruskal-Wallis does not find a statistical difference (H =  1.3195; df = 2; $p$ = 0.517). In the same way, the criterion ``Connection to previous knowledge'' did not show a statistically significant difference (H = 0.3882; df = 2; $p$ = 0.8236).

In the ``Final average'' aspect, the Kruskal-Wallis test identified statistically significant differences (H = 7.0449; $p$ = 0.0295). Dunn's test with Bonferroni correction revealed that ChatGPT differed significantly from DeepSeek ($p_{bonferroni}$ = 0.0238) but not from Gemini ($p_{bonferroni}$ = 0.5599). DeepSeek and Gemini did not show a statistically significant difference ($p_{bonferroni}$ = 0.5470). Table \ref{tab:dun-explanations-average} in \ref{app:statistical} presents these values. 

In the ``Clarity, consistency, and ease'' aspect, the Kruskal-Wallis test showed a statistically significant difference (H = 7.6037; df = 2; p = 0.0223). Dunn's test, with Bonferroni correction, indicated that ChatGPT showed differences significantly from DeepSeek ($p_{bonferroni}$ = 0.0262), but did not differ from Gemini ($p_{bonferroni}$ = 1.000). Gemini and DeepSeek ($p_{bonferroni}$ = 0.1215) also did not show statistically significant differences for this criterion. Detailed values are presented in Table \ref{tab:dun-explanations-clarity} in \ref{app:statistical}.

For the ``Critical parts focus'' criterion, the Kruskal-Wallis detected a statistically significant difference (H = 53.9816; df = 2; $p < 0.001$) among the models. Dunn's test revealed a significant difference between ChatGPT and DeepSeek ($p_{bonferroni} < 0.001$) and between DeepSeek and Gemini ($p_{bonferroni} < 0.001$). However, Gemini and ChatGPT did not show a significant difference ($p_{bonferroni}$ = 0.0985). For a detailed comparison of these results, see Table \ref{tab:dun-explanations-focus} in \ref{app:statistical}.

The Kruskal-Wallis test for the criterion of ``Level adaptation'' identified a statistically significant difference (H = 13.8414; df = 2; $p$ = 0.0010). Dunn's test with Bonferroni correction showed significant differences between ChatGPT and DeepSeek ($p_{bonferroni}$ = 0.0057) and between DeepSeek and Gemini ($p_{bonferroni}$ = 0.0026). However, ChatGPT and Gemini ($p_{bonferroni}$ = 1.0000) did not show a significant difference. Refer to Table \ref{tab:dun-explanations-level} for the complete statistical results.

For the ``Usefulness'' criterion, the Kruskal-Wallis detected a statistically significant difference (H = 26.8662; df = 2; $p < 0.001$). Dunn's test revealed significant differences between ChatGPT and DeepSeek ($p_{bonferroni}$ = 0.0005) and between ChatGPT and Gemini ($p_{bonferroni} < 0.001$). DeepSeek and Gemini ($p_{bonferroni}$ = 0.7126) did not differ significantly. These values are presented in Table \ref{tab:dun-explanations-usefulness} in \ref{app:statistical}.

\begin{table}[htbp]
\centering
\caption{Kruskal–Wallis Test Results for Models in Explanations and Analogies Approach (df = 2; n = 150)}
\label{tab:kruskal-explanations}
\begin{tabular}{lrr}
\hline
Criterion & H statistic & p-value \\
\hline
Final average$^{*}$ & 7.0449 & 0.0295 \\
Clarity, consistency, and ease$^{*}$ & 7.6037 & 0.0223 \\
Critical parts focus$^{*}$ & 53.9816 & $ < 0.001$ \\
Correctness &  1.3195 & 0.5170 \\
Level adaptation$^{*}$ & 13.8414 & 0.0010 \\
Usefulness$^{*}$ & 26.8662 & $ < 0.001$ \\
Previous knowledge connection & 0.3882 & 0.8236 \\
\hline

\end{tabular}

\smallskip
\textit{$*$ Variables that exhibited a statistically significant difference.}
\end{table}

\begin{figure}[H]
\centering
\includegraphics[width=0.70\textwidth]{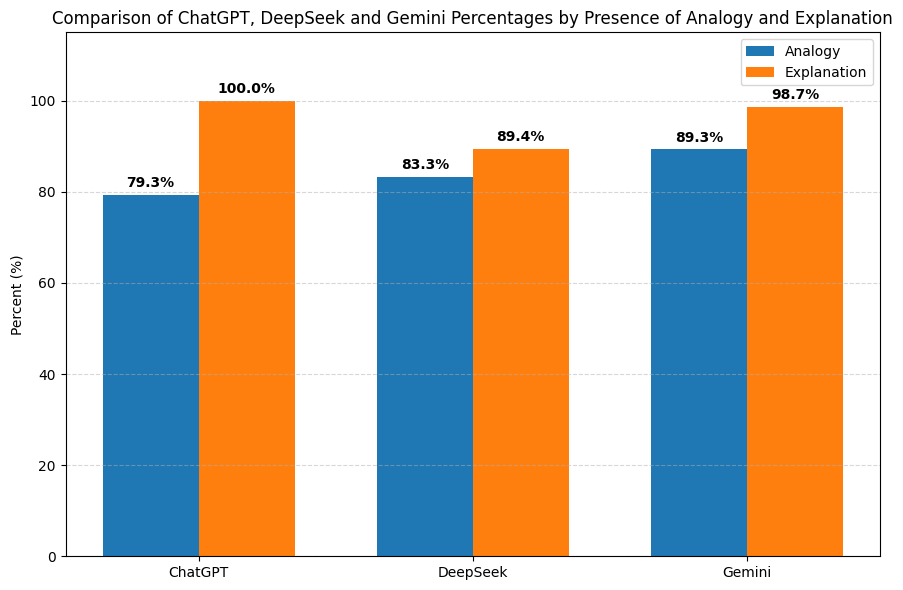}
\caption{Comparison of ChatGPT, DeepSeek, and Gemini Percentages by Presence of Analogy and Explanation}\label{fig:analogy-explanation}
\end{figure}

To further the analysis, we examined whether explanations and analogies were used, considering them separately, since both represent fundamental aspects of the pedagogical approach under evaluation. Thus, the judges responded to the presence of each element in each assessment (Figure \ref{fig:analogy-explanation}). ChatGPT presented analogies in 79.3\% of the cases and explanations in 100\%. Gemini presented analogies in 89.3\% and explanations in 98.7\%. DeepSeek presented analogies in 83.3\% and explanations in 89.4\% of the evaluations. These results suggest that the models presented more explanations than analogies in the evaluations, indicating that implementing the analogy strategy may be more challenging for LLMs.

\begin{figure}[H]
\centering
\includegraphics[width=0.70\textwidth]{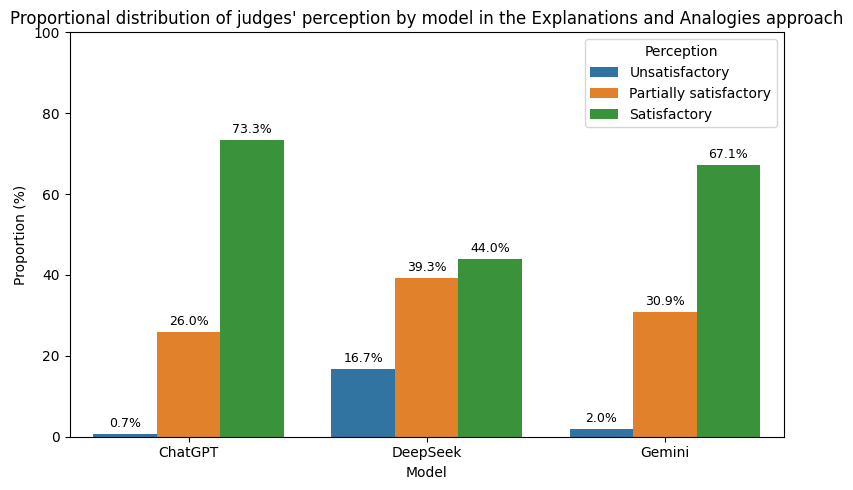}
\caption{``Judges' perception'' of Explanations and Analogies approach}\label{fig:explanations-decision}
\end{figure}

Figure \ref{fig:explanations-decision} presents the results of the ``Judge perception'' criterion. As can be seen, DeepSeek (44\%``satisfactory''; 39.3\% ``partially satisfactory''; 16.7\% ``unsatisfactory'') received less satisfactory evaluations compared to both Gemini (67.1\%``satisfactory''; 30.9\% ``partially satisfactory''; 2\% ``unsatisfactory'') and ChatGPT (73.3\%``satisfactory''; 26\% ``partially satisfactory''; 0.7\% ``unsatisfactory'').
The Chi-squared test revealed statistically significant differences ($\chi^2 = 52.4613$;  $p < 0.001$). Following the Bonferroni correction test, significant differences were found between ChatGPT and DeepSeek ($p < 0.001$; $p_{Bonferroni} < 0.001$) and between Gemini and DeepSeek ($p < 0.001$; $p_{Bonferroni} < 0.001$). In contrast, ChatGPT and Gemini ($p = 0.3589$; $p_{Bonferroni} = 1.0000$) did not show any significant differences. The complete results are presented in Table \ref{tab:dun-explanations-perception} in \ref{app:statistical}.

\begin{figure}[H]
\centering
\includegraphics[width=0.70\textwidth]{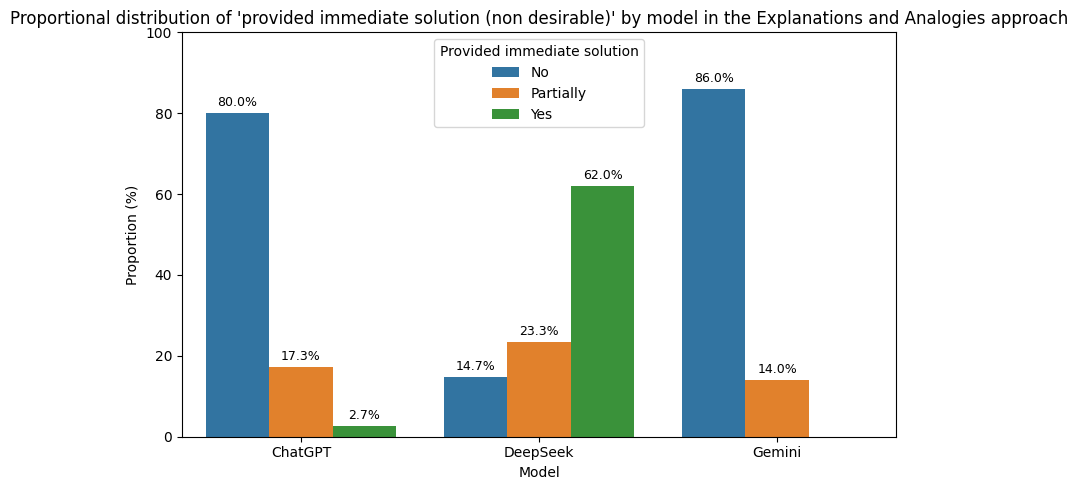}
\caption{``Provided immediate solution (non desirable)'' criterion of Explanations and Analogies approach}\label{fig:explanations-answer}
\end{figure}

The results of the analysis of the models’ tendency to provide immediate solutions (an undesirable behavior) under this approach are presented in Figure \ref{fig:explanations-answer}. 
DeepSeek was the model that provided the most immediate solutions, with 62\% of its responses classified as ``yes'', 23.3\% as ``partially'', and only 14.7\% as ``no''. ChatGPT (2.7\% ``yes''; 17.3\% ``partially''; 80\% ``no'') and Gemini (0\% ``yes''; 14\% ``partially''; 86\% ``no''), on the other hand, showed similar values and a more desirable behavior, refraining from providing immediate solutions and thus helping students to achieve their own answers. The Chi-square test found a significant difference ($\chi^2 = 252.658$; $p < 0.001$). In the Bonferroni correction test, similar to the previous criterion, statistically significant differences were found between ChatGPT and DeepSeek ($p < 0.001$; $p_{bonferroni} < 0.001$) and between Gemini and DeepSeek ($p < 0.001$; $p_{bonferroni} < 0.001$). However, ChatGPT and Gemini ($p = 0.0882$; $p_{bonferroni} = 0.2645$) did not show a significant difference. Table \ref{tab:dun-explanations-solution} in \ref{app:statistical} presents these results.

\begin{figure}[H]
\centering
\includegraphics[width=0.80\textwidth]{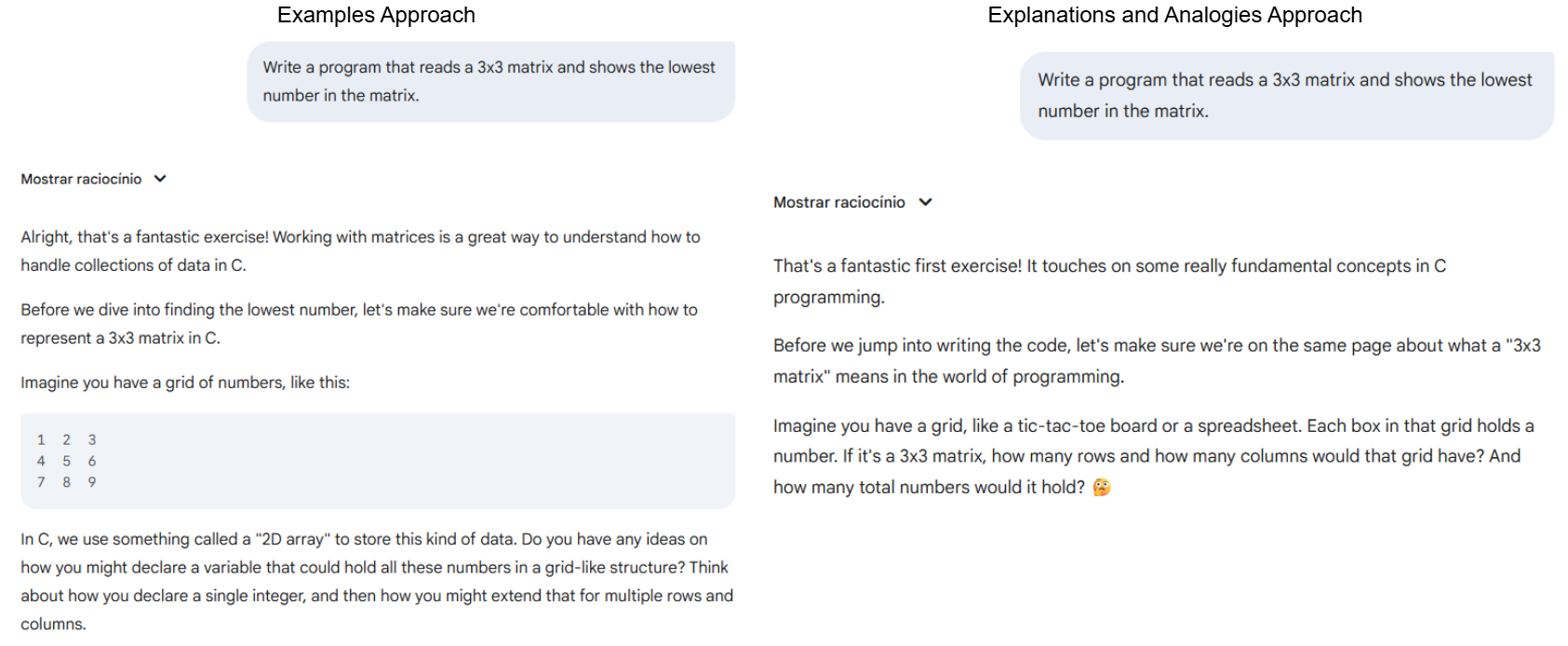}
\caption{Answers of Gemini 2.5 flash in Examples Approach and Explanations and Analogies Approach for the same exercise}\label{fig:gemini-examples-explanations-4.3}
\end{figure}

In addition to the obtained results, the judges observed certain behaviors of the models during the interactions. Notably, they highlighted the similarity in response styles across all models when using the Examples and Explanations and Analogies approaches. For instance, Figure \ref{fig:gemini-examples-explanations-4.3} shows that Gemini produced very similar responses in both approaches. 

\begin{figure}[H]
\centering
\includegraphics[width=0.80\textwidth]{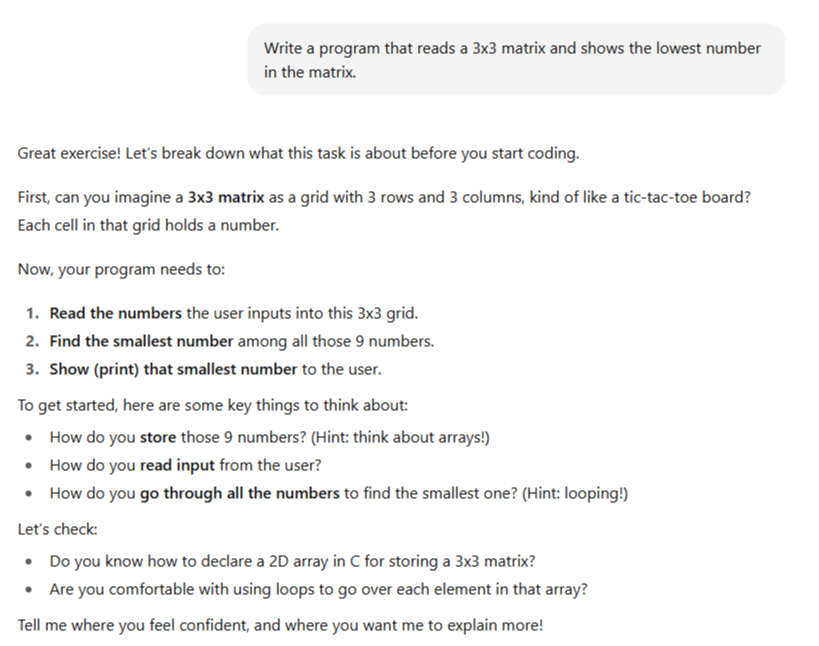}
\caption{ChatGPT 4.1-mini in Explanations and Analogies Approach}\label{fig:chat-explanations-4.3}
\end{figure}

Another observed behavior was related to the length of the responses given; in both approaches, the three models generally provided extensive answers. Figure \ref{fig:chat-explanations-4.3} presents a long answer given by ChatGPT; as can be observed, all resolution steps are listed, and many questions are asked in the same answer. 

\subsection{Socratic Method Approach}

The results obtained in the evaluation process of the Socratic Method are presented in Table \ref{tab:socratic}. Taking into account the ``Final average'' criterion, it is possible to conclude that ChatGPT (4.631) generally performed best in the evaluations, followed by Gemini (4.427) and DeepSeek (4.051). 
In the ``Initial question'' criterion, ChatGPT (4.620) scored significantly higher than Gemini (4.413). This indicates that, in the judges' evaluations, ChatGPT was more effective in providing an initial question to the student, compared to Gemini. However, DeepSeek (4.587) had average scores very similar to both models. 
In the criterion of ``Counterexamples'', DeepSeek (3.313) was evaluated with the smallest scores compared to ChatGPT (4.413) and Gemini (4.160) in this criterion. Thus, it can be observed that DeepSeek did not present sufficient counterexamples to test the validity of the answers provided by the judge.

In the ``Questions only'' aspect, ChatGPT (4.800) and Gemini (4.653) were considered more satisfactory compared to DeepSeek (4.200), which, in turn, achieved the lowest ratings for this criterion.
In the ``Well-formulated'' criterion, DeepSeek (4.193) obtained lower scores compared to ChatGPT (4.647) and Gemini (M = 4.500), indicating that, in comparison, its questions were less effective at encouraging investigation. 
In the ``Critical thinking promotion'' aspect, DeepSeek (3.960) presented the lowest average, Gemini (4.407) was second, and the highest average was with ChatGPT (4.673). 

\begin{table}[htbp]
\centering
\caption{Descriptive statistics of the criteria evaluated for ChatGPT, DeepSeek, and Gemini in the Socratic Method Approach}
\label{tab:socratic}
\renewcommand{\arraystretch}{1.2}
\setlength{\tabcolsep}{3pt}
\begin{tabular}{lccccccccc}
\hline
\multirow{2}{*}{\textbf{Criteria}} & 
\multicolumn{3}{c}{\textbf{ChatGPT}} & 
\multicolumn{3}{c}{\textbf{DeepSeek}} & 
\multicolumn{3}{c}{\textbf{Gemini}} \\ 
& \textbf{M} & \textbf{Md} & \textbf{SD}
& \textbf{M} & \textbf{Md} & \textbf{SD}
& \textbf{M} & \textbf{Md} & \textbf{SD} \\ \hline
Final average       & 4.631 & 4.600       & 0.423   & 4.051   & 4.200   & 0.839 & 4.427     & 4.600   & 0.524 \\
Initial question    & 4.620 & 5.000       & 0.825   & 4.587   & 5.000   & 0.813 & 4.413     & 5.000   & 0.845  \\
Counterexamples     & 4.413 & 5.000       & 0.906   & 3.313   & 4.000   & 1.777 & 4.160     & 5.000   & 1.124 \\
Questions only      & 4.800 & 5.000       & 0.543   & 4.200   & 5.000   & 1.074 & 4.653     & 5.000   & 0.655 \\
Well-formulated \\
questions           & 4.647 & 5.000       & 0.761   & 4.193   & 4.500   & 1.079 & 4.500     & 5.000   & 0.857  \\
Critical thinking \\
promotion           & 4.673 & 5.000       & 0.709   & 3.960   & 4.000   & 1.092 & 4.407     & 5.000   & 0.875  \\
\hline
\end{tabular}

\smallskip
\textit{M = Mean; Md = Median; SD = Standard Deviation.}
\end{table}

\begin{figure}[H]
\centering
\includegraphics[width=1\textwidth]{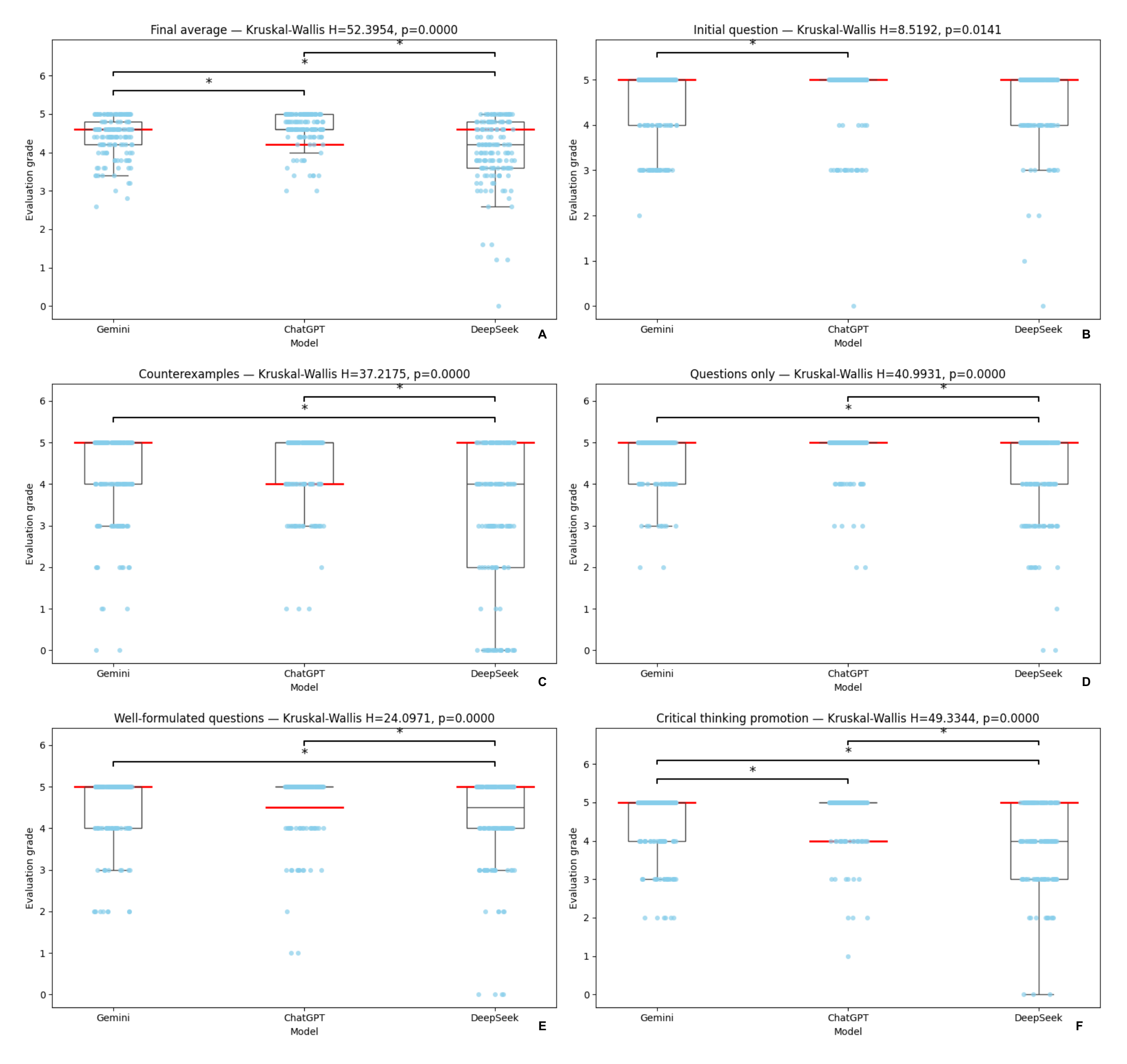}
\caption{Distribution of evaluation grades for Socratic Method approach for the criteria: (A) ``Final average''; (B) ``Initial question''; (C) ``Counterexamples''; (D) ``Questions only''; (E) ``Well-formulated questions'' criterion; (F) ``Critical thinking promotion''. Blue dots represent individual observations; boxes indicate the data distribution (quartiles), and the red line marks the median of each group — the value dividing the dataset into two equal halves and serving as a robust measure of central tendency. Black lines with asterisks denote statistically significant differences between models, identified by Dunn’s post hoc test ($p < 0.05$) following the Kruskal–Wallis test.}\label{fig:socratic}
\end{figure}

Statistical analysis outcomes are presented in Table \ref{tab:kruskal-socratic}, with the corresponding distributional patterns illustrated in Figure \ref{fig:socratic}.
The Kruskal-Wallis test of the ``Final average'' criterion identified a statistically significant difference (H = 52.3954; $p < 0.001$) between the final average scores of the models. As a result, Dunn's test revealed differences between ChatGPT and DeepSeek ($p_{bonferroni} < 0.001$), DeepSeek and Gemini ($p_{bonferroni} < 0.001$), and ChatGPT and Gemini ($p_bonferroni$ = 0.0017). These results are presented in Table \ref{tab:dun-socratic-average} in \ref{app:statistical}.

For the criterion of the ``Initial question'', the Kruskal-Wallis test identified a statistically significant difference between the models (H = 8.5192; $p$ = 0.0141). Dunn's test revealed a statistically significant difference between ChatGPT and Gemini ($p_{bonferroni}$ = 0.0124), but not between ChatGPT and DeepSeek ($p_{bonferroni}$ = 0.9967) or DeepSeek and Gemini ($p_{bonferroni}$ = 0.1726). Table \ref{tab:dun-socratic-initial} in \ref{app:statistical} presents the detailed results.

In the ``Counterexamples'' criterion, the Kruskal-Wallis test showed a statistical difference between models (H = 37.2175; $p < 0.001$). In Dunn's test, the comparison between ChatGPT and DeepSeek ($p < 0.001$) and DeepSeek and Gemini ($p < 0.001$) presented a significant difference. In contrast, ChatGPT and Gemini ($p$ = 0.2035) did not present differences. These results are presented in Table \ref{tab:dun-socratic-counterexamples} in \ref{app:statistical}.

In the Kruskal-Wallis test for the criterion ``Question only'', the p-value showed that there is a significant statistical difference (H = 40.9931; $p < 0.001$). In the Dunn test, ChatGPT and DeepSeek ($_{bonferroni} < 0.001$), and Gemini and DeepSeek ($p_bonferroni < 0.001$) found a statistically significant difference, however, ChatGPT and Gemini ($p_{bonferroni}$ = 0.1224) did not. Further statistical details can be found in Table \ref{tab:dun-socratic-questions}, located in \ref{app:statistical}.

The same applies for the criterion ``Well-formulated questions'', the Kruskal-Wallis test found a significant statistical difference (H = 40.9931; $p < 0.001$). In Dunn's test, ChatGPT and DeepSeek ($p_{bonferroni} < 0.0001$), and Gemini and DeepSeek ($p_{onferroni}$ = 0.0045) showed differences; however, ChatGPT and Gemini ($p_{bonferroni}$ = 0.2957) did not show differences in their grades. Table \ref{tab:dun-socratic-wellformulated} in \ref{app:statistical} shows the detailed values.

As a final point, the Kruskal-Wallis test found a significant statistical difference (H = 49.3344; $p < 0.001$) in the ``Critical thinking promotion'' criterion. The Dunn test encountered a difference between all pairs of LLMs: ChatGPT and DeepSeek ($p_{bonferroni} < 0.001$), ChatGPT and Gemini ($p_{bonferroni}$ = 0.0175) and Gemini and DeepSeek ($p_{bonferroni} = 0.0001$). The full set of results is available in Table \ref{tab:dun-socratic-critical} (\ref{app:statistical}).

\begin{table}[htbp]
\centering
\caption{Kruskal–Wallis Test Results for Models in Socratic Method Approach (df = 2; n = 150)}
\label{tab:kruskal-socratic}
\begin{tabular}{lrr}
\hline
Criterion & H statistic & p-value \\
\hline
Final average$^{*}$ & 52.3954  & $<0.001$\\
Initial question$^{*}$ & 8.5192 & 0.0141 \\
Counterexamples$^{*}$ & 37.2175 & $<0.001$ \\
Questions only$^{*}$ & 40.9931 & $<0.001$ \\
Well-formulated questions$^{*}$ & 24.0971 & $<0.001$  \\
Critical thinking promotion$^{*}$ & 49.3344 & $<0.001$ \\
\hline
\end{tabular}

\smallskip
\textit{$^{*}$ variables that exhibited a statistically significant difference.}
\end{table}

\begin{figure}[H]
\centering
\includegraphics[width=1.0\textwidth]{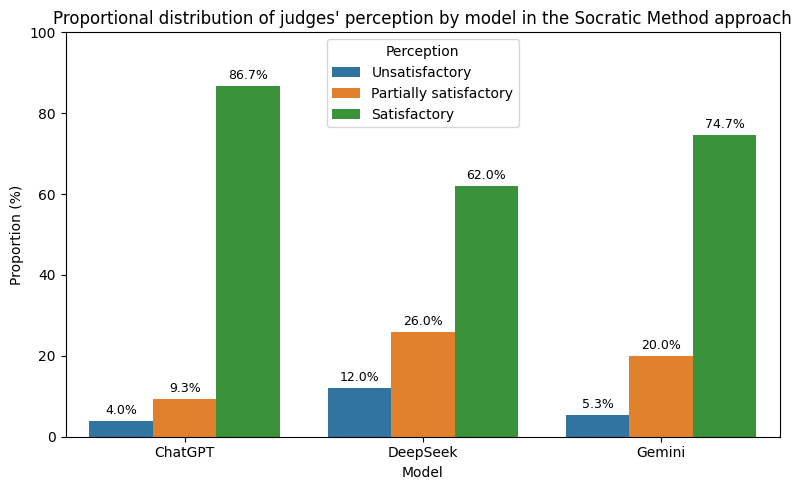}
\caption{``Judges' perception'' of Socratic Method approach}\label{fig:socratic-final}
\end{figure}

To analyze the ``Judges' perception'' of each model, Figure \ref{fig:socratic-final} presents the results obtained. 
DeepSeek received evaluations of 62.0\% as ``satisfactory'', 26\% as ``partially satisfactory'', and 12\% as ``unsatisfactory''. Gemini was evaluated as ``satisfactory'' in 74.7\% of cases, ``partially satisfactory'' in 20\%, and ``unsatisfactory'' in 5.3\%. 
ChatGPT, outperforming DeepSeek, received satisfactory evaluations in 86.7\% of cases, ``partially satisfactory'' in 9.3\%, and only 4\% as ``unsatisfactory''. 
The Chi-square test detected a statistically significant difference ($\chi^2$ = 25.471, $p < 0.001$). 
In the Chi-square test with Bonferroni correction, only ChatGPT and DeepSeek ($p < 0.001$; $p_{bonferroni} < 0.001$) showed a statistically significant difference, while comparisons between ChatGPT and Gemini ($p = 0.0242$; $p_{bonferroni} = 0.0726$), and between DeepSeek and Gemini ($p = 0.0337$; $p_{bonferroni} = 0.1011$), indicated no statistically significant differences. Table \ref{tab:dun-socratic-perception} in \ref{app:statistical} presents the complete results.

\begin{figure}[H]
\centering
\includegraphics[width=1.0\textwidth]{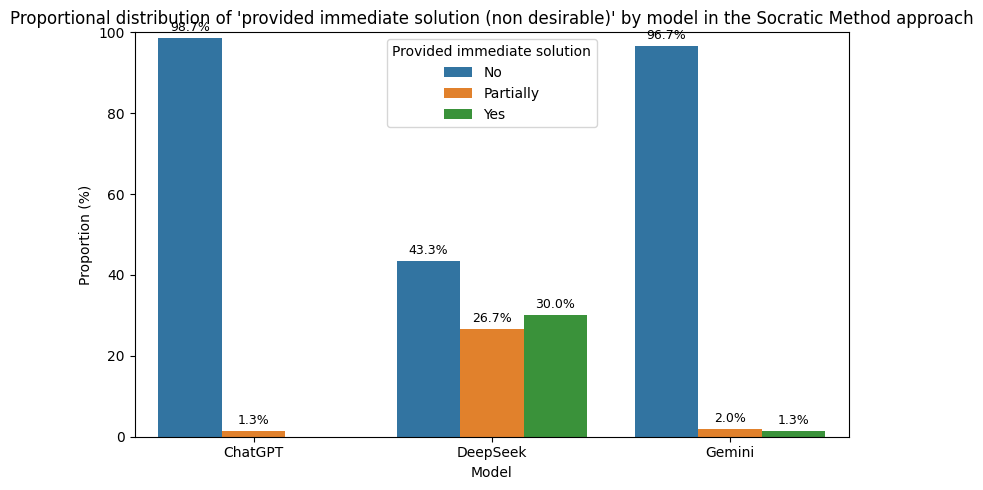}
\caption{``Provided immediate solution (non desirable)'' criterion of Socratic Method approach}\label{fig:socratic-answer}
\end{figure}

In the behavior of provide immediate solutions responses, as shown in Figure \ref{fig:socratic-answer}, DeepSeek (30\% ``yes''; 26.7\% ``partially''; 43.3\% ``no'') provided more immediate solutions during the evaluations compared to the other models. ChatGPT (0\% ``yes''; 1.3\% ``partially''; 98.7\% ``no'') and Gemini (1.3\% ``yes''; 2\% ``partially''; 96.7\% ``no'') had similar evaluations and exhibited a more desirable behavior, rarely providing immediate solutions. 
In the Chi-square test, performed ($\chi^2$ = 182.189, $p < 0.001$), it was revealed a statistically significant difference. In the Chi-square test with Bonferroni correction, statistically significant differences were found between ChatGPT and DeepSeek ($p < 0.001$; $p_{bonferroni} < 0.001$), and between Gemini and DeepSeek ($p < 0.001$; $p_{bonferroni} < 0.001$), while ChatGPT and Gemini ($p < 0.3278$; $p_{bonferroni} < 0.9834$) did not show significant differences. Detailed results are reported in Table \ref{tab:dun-socratic-solution}, provided in \ref{app:statistical}.

\begin{figure}[H]
\centering
\includegraphics[width=0.80\textwidth]{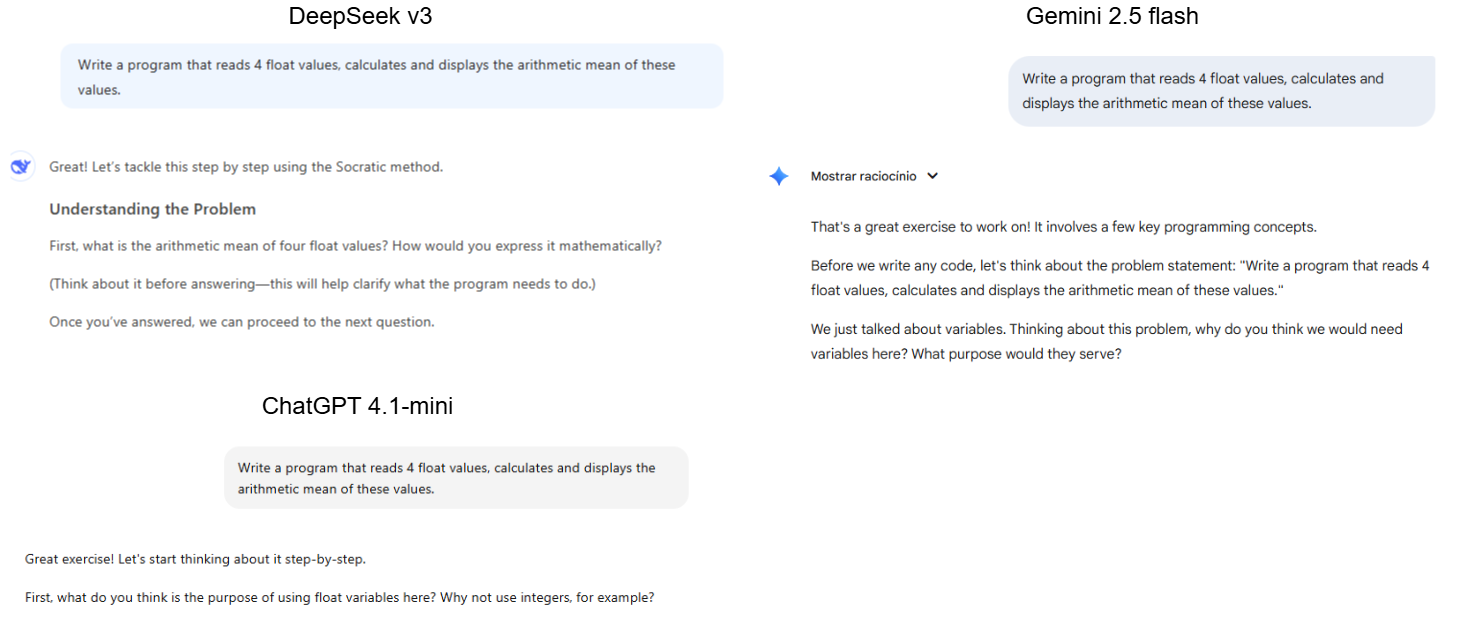}
\caption{The three LLMs in Socratic Method Approach}\label{fig:socratic-1.2}
\end{figure}

According to the judges' perception, the models' behavior in this approach was different from the others. They provided more direct answers and effectively led the conversation process with questions.
As shown in Figure \ref{fig:socratic-1.2}, the responses of the three models were verified during one of the evaluations of the Socratic method. All models provided a direct initial question in accordance with the initial prompt's requirements.

\begin{figure}[H]
\centering
\includegraphics[width=0.80\textwidth]{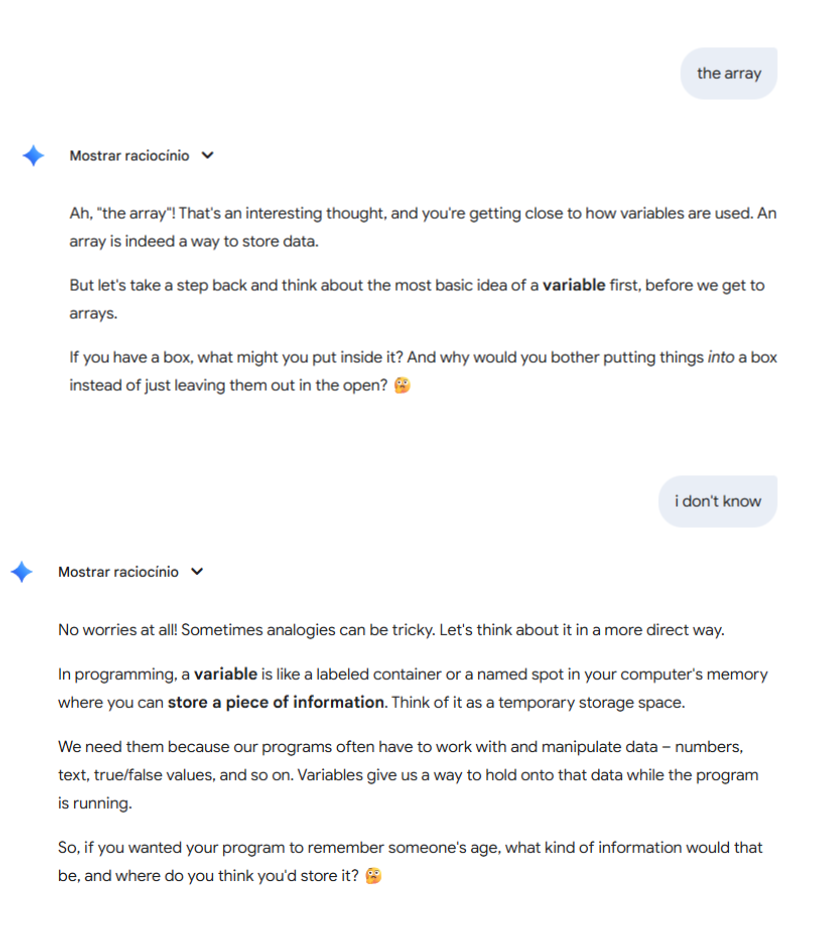}
\caption{Gemini 2.5 flash in Socratic Method Approach}\label{fig:gemini-socratic-4.2}
\end{figure}

Another observation about Socratic Method, in Gemini, the models began to lead with very fundamental concepts. For example, even if the question was about arrays, Gemini started the conversation about variables, often not reaching the aspects of the target exercise for that interaction within the limit established in the evaluation protocol of 5-10 interactions. 
Although this behavior is interesting because it starts with foundational concepts, in practice, it can make interactions tedious for the student if it is not aligned with their level. An example is presented in Figure \ref{fig:gemini-socratic-4.2}, in which the exercise was about arrays, and Gemini persisted in variable concepts.

\section{Limitations}\label{sec:lim}
Several limitations of this study should be considered. The first limitation to highlight is the origin of the selected LLMs, as ChatGPT and Gemini are Western, while DeepSeek is Eastern. This may have affected the evaluation due to cultural biases inherent in the assessment process, since the judges are also Western. Although the same criteria were used for all models to mitigate this limitation, it does not completely eliminate each judge's individual bias. However, a detailed analysis of these aspects is beyond the scope of this paper.

Another limitation to note is the changes that the models underwent during the testing period. For example, during testing, ChatGPT released the 4.1 model and removed the 4o-mini version. In response, the tests were reworked on the new model so that all judges considered the same version during their evaluations. In this regard, the test battery's comprehensiveness may have made it potentially stressful for each judge, thereby influencing their evaluations.

Furthermore, the experimental design relied exclusively on expert judges simulating novice learners. While necessary for a controlled evaluation, this simulation may not fully capture the specific misconceptions or linguistic patterns of actual learners, thereby limiting the ecological validity of the findings regarding student interaction. 

Additionally, the judges interacted with the models using their native web interfaces, meaning the evaluation was not blinded. We acknowledge that knowing the model's identity could introduce brand bias, potentially influencing the judges' assessments based on their prior expectations.

Lastly, since the goal was to analyze the behavior of the model with beginner students in the C programming language, the exercise selection was consistent and accounted for this difficulty level. However, this may have limited the range of possible responses. In other words, it might not have allowed the LLMs to provide more examples, explanations, and analogies, or to ask questions that would foster more critical thinking.

\section{Conclusion and Future Work}\label{sec:con}

This study addressed the research question: Are LLMs capable of teaching using specific pedagogical approaches? The article presented a detailed evaluation protocol and reported the results of the analyses, providing insights into how the LLMs ChatGPT, DeepSeek, and Gemini performed across the pedagogical strategies of Examples, Explanations and Analogies, and the Socratic Method.

When evaluating the three strategies, we observed that the models' responses were influenced by the initial instructions and by their own knowledge of the pedagogical approaches. In the Examples and Explanations and Analogies strategies, ChatGPT and Gemini exhibited very similar behaviors, frequently providing explanations even when examples were expected. These patterns highlight the models’ tendencies to generalize across closely related pedagogical strategies. 

In contrast, performance in the Socratic Method was more variable across the models. The interactions revealed that the models' ability to engage in guided questioning was highly dependent on the initial prompt and the formulation of the questions. Although ChatGPT and Gemini generally followed the intended Socratic structure, DeepSeek often struggled to maintain a consistent line of inquiry, occasionally providing direct answers rather than prompting reflection. These findings suggest that LLMs can adopt Socratic-like teaching behaviors, but their effectiveness is highly sensitive to prompt design and alignment of strategies.

To facilitate visualization of the results, we integrated the evaluations with the final analyzes and expressed them as star ratings. 

\begin{figure}[H]
\centering
\includegraphics[width=0.80\textwidth]{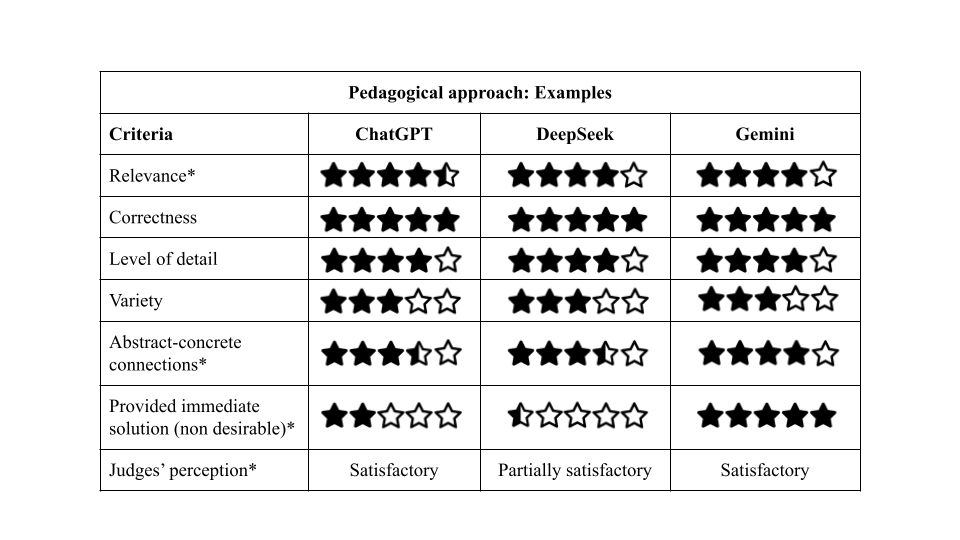}
\caption{Evaluation of Pedagogical Skills Using the Examples approach. Asterisks indicate variables that showed a statistically significant difference in the statistical analysis. }\label{fig:stars-example}
\end{figure}

The final ranking of the models can be verified in our analysis, as shown in Figure \ref{fig:stars-example}, within the pedagogical approach of Examples. The criteria marked with asterisks highlight those that showed statistically significant differences in the analysis conducted. In the ``Relevance'' criterion, ChatGPT outperformed DeepSeek and Gemini, which showed similar results. For ``Abstract-concrete concepts'', Gemini achieved the highest scores. In the ``Correctness'' and ``Level of Detail'' criteria, all three models performed satisfactorily, while ``Variety'' received the lowest scores across all models. Regarding the ``Providing immediate solutions'' criterion, Gemini obtained the best results. Overall, according to the judges’ perception, DeepSeek was rated ``partially satisfactory'', while ChatGPT and Gemini were considered ``satisfactory''.

\begin{figure}[H]
\centering
\includegraphics[width=0.80\textwidth]{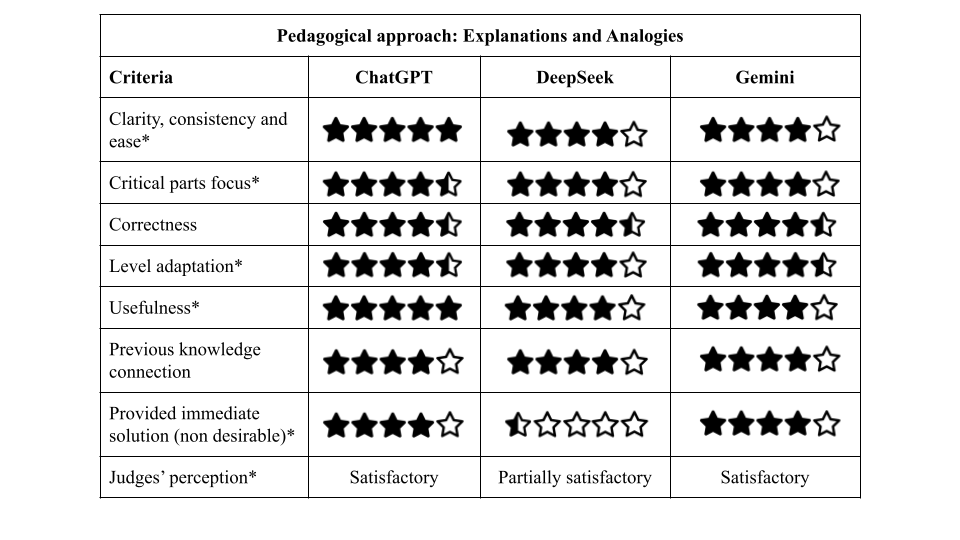}
\caption{Evaluation of Pedagogical Skills Using the Explanations and Analogies approach. Asterisks indicate variables that showed a statistically significant difference in the statistical analysis.}\label{fig:stars-explanations}
\end{figure}

The final ranking of the models for the Explanations and Analogies pedagogical approach is shown in Figure \ref{fig:stars-explanations}. ChatGPT outperformed DeepSeek and Gemini in the ``Clarity, consistency and ease'', ``Critical parts focus'', and ``Usefulness'' criteria, receiving the highest scores. All three models performed satisfactorily on the ``Correctness'' criterion. DeepSeek obtained the lowest scores on the ``Level adaptation'' and ``Provided immediate solutions'' criteria. Overall, based on the Judges’ perception, ChatGPT and Gemini were mostly rated as ``satisfactory'', whereas DeepSeek was considered ``partially satisfactory''.

\begin{figure}[H]
\centering
\includegraphics[width=0.80\textwidth]{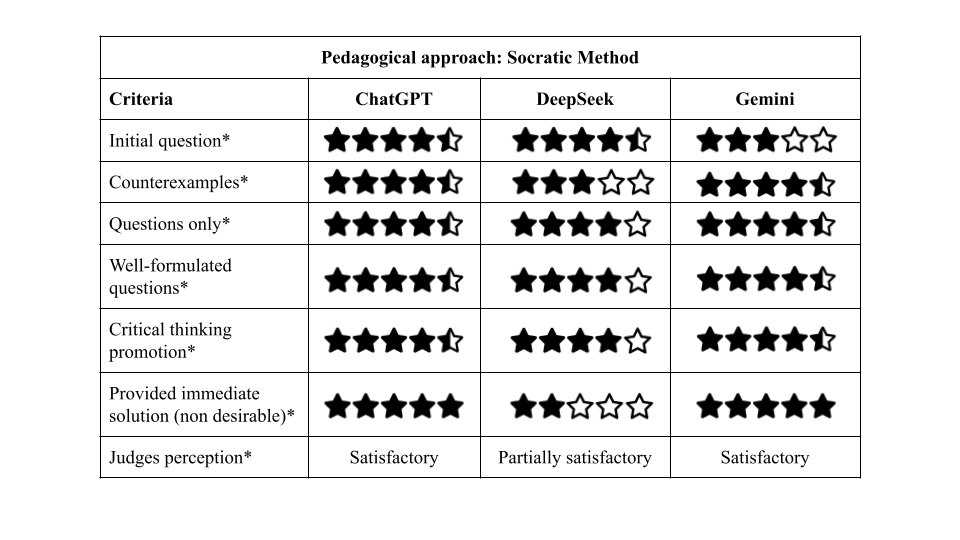}
\caption{Evaluation of Pedagogical Skills Using the Socratic Method approach. Asterisks indicate variables that showed a statistically significant difference in the statistical analysis.}\label{fig:stars-socratic}
\end{figure}

Figure \ref{fig:stars-socratic} displays the final ranking of the models for the Socratic Method pedagogical approach. In the ``Initial question'' criterion, Gemini received the lowest scores, whereas ChatGPT and DeepSeek performed more satisfactorily. For the ``Counterexamples'', ``Questions only'', ``Well-formulated questions'', ``Critical thinking promotion'', and ``Provided immediate solutions'' criteria, DeepSeek obtained the lowest scores, while ChatGPT and Gemini performed better. Overall, consistent with the other two pedagogical approaches, the Judges’ perception rated ChatGPT and Gemini as ``satisfactory'', while DeepSeek was considered ``partially satisfactory''.

The evaluation results indicate that ChatGPT and Gemini performed more satisfactorily as teacher agents, particularly when using the Socratic Method prompt. This approach stood out because the models demonstrated greater fluency in adhering to the initial prompt, posing more direct and focused questions to guide the step-by-step resolution of the exercises. In contrast, DeepSeek showed lower performance, highlighting limitations that suggest the need for prompt adaptation or the exploration of alternative pedagogical strategies.

Based on these findings, this article contributes to the community in three main aspects: (i) an LLM evaluation protocol based on pedagogical approaches that can be extended to other models, (ii) a comparative analysis of pedagogical strategies for LLM teacher agents, and (iii) empirical results from a comprehensive evaluation of ChatGPT, DeepSeek, and Gemini.

For future work, we plan to include additional models and updated versions of the current LLMs. Moreover, we aim to extend the analysis to include inter-rater reliability among judges, addressing the methodological challenges imposed by the dynamic nature of LLM interactions. Additionally, we will explore a deep analysis considering the level of complexity of exercises and real scenarios with students in the classroom, including evaluating the learning gains through pre- and post-test assessments.

Furthermore, the protocol is expected to be applied in the study modes provided by ChatGPT \citep{chat-study} and Gemini \citep{gemini-study}, as well as in resource-limited contexts, as envisioned in AIED Unplugged, to investigate its applicability in scenarios where students have limited access to digital tools and by using Small Language Models (SLMs). 

\section{Ethics Statement}
Given that this work investigates the pedagogical capabilities of proprietary LLMs, it is crucial to address the ethical implications of their deployment. The evaluated models are computationally intensive and reliant on cloud infrastructure, requiring high-bandwidth internet connectivity and, in many cases, paid subscriptions. These requirements pose significant adoption barriers in resource-limited contexts. Consequently, developing educational resources that rely exclusively on these systems carries the risk of exacerbating the digital divide. While this study validated the baseline capabilities of current-state-of-the-art LLMs, our future work aims to address issues of bias and accessibility. Specifically, we intend to investigate Small Language Models (SLMs) that can be deployed locally, ensuring that the benefits of AI in Education are equitably accessible across diverse socioeconomic contexts.
Finally, we clarify that the deployment of these models in educational contexts is investigated not as a replacement for human teachers but as a form of AI augmentation. These models are positioned as supportive agents designed to offload repetitive tasks, thereby enabling teachers to focus on complex pedagogical activities and classroom dynamics.

\section{Declaration of generative AI and AI-assisted technologies in the manuscript preparation process}
During the preparation of this work the authors used ChatGPT and Gemini in order to support text revision and language corrections.
After using this tool/service, the authors reviewed and edited the content as needed and take full responsibility for the content of the published article.

\appendix
%TC:ignore
\section{List of Exercises}\label{app:exercises}
Selected exercises from \cite{backes2013linguagem}:
\begin{itemize}
    \item Task 1.1 - operators - Write a program that reads an integer number and returns its predecessor and successor.
    \item Task 1.2 - operators - Write a program that reads 4 float values, calculates and displays the arithmetic mean of these values.
    \item Task 1.3 - operators - Write a program that calculates a person's year of birth from their age and the current year.
    \item Task 1.4 - operators - Write a program that reads a dollar value and the real exchange rate. Then, display the corresponding real value.
    \item Task 1.5 - operators -Write a program that reads a velocity in km/h (kilometers per hour) and shows it converted to m/s (meters per second). The conversion formula is M=K/3, where K is the velocity in km/h and M is the velocity in m/s.
    \item Task 2.1 - condition - Write a program that reads two integer numbers and shows which one is the largest.
    \item Task 2.2 - condition - Write a program that reads an integer number and checks if it is even or odd.
    \item Task 2.3 - condition - Write a program that reads a number and if it is positive, calculates and shows: a) the square of the number, b) the square root of the number.
    \item Task 2.4 - condition - Write a program that checks if an integer number is divisible by 3 or 5, but not simultaneously by both.
    \item Task 2.5 - condition - Write a program that reads 3 numbers and shows them in ascending order.
    \item Task 3.1 - loops - Write a program that reads a positive integer number N and displays all the natural numbers from 0 to N in ascending order.
    \item Task 3.2 - loops - Write a program that reads an integer number N and then shows the first N odd natural numbers
    \item Task 3.3 - loops - Write a program that reads 10 integer numbers and shows their mean.
    \item Task 3.4 - loops - Write a program that receives an integer number greater than 1 and checks if the number is prime or not.
    \item Task 3.5 - loops - Write a program that reads a set of numbers, shows the largest one and how many times it was read. The quantity of numbers should be provided by the user.
    \item Task 4.1 - array - Write a program that reads six values and then shows the values read.
    \item Task 4.2 - array - Write a program that reads an array with 8 positions. Then, reads two values X and Y corresponding to two array positions. Your program should show the sum of the values at X and Y positions.
    \item Task 4.3 - array - Write a program that reads a 3x3 matrix and shows the lowest number in the matrix.
    \item Task 4.4 - array - Write a program that reads a 3x3 matrix and then shows the sum of the numbers in the main diagonal.
    \item Task 4.5 - array - Write a program that reads a 5x5 matrix. Calculate and print the sum of the elements in this matrix that are above the main diagonal.
    \item Task 5.1 - function - Write a function that receives two parameters and returns the largest one.
    \item Task 5.2 - function - Write a function that receives an integer number from 1 to 12 and prints the month corresponding to the typed number.
    \item Task 5.3 - function - Write a function that receives an array of 10 elements and returns its sum.
    \item Task 5.4 - function - Write a function that receives a string and convert all characters to uppercase.
    \item Task 5.5 - function - Write a function that receives the weight (kg) and the heigh (meters) of a person. Calculate and return the BMI (Body Mass Index) from this person: BMI= weight/(height*height)
\end{itemize}

\section{Prompts for Pedagogical Approaches}\label{app:prompts}
For the Example approach, the following final prompt was used: 
\begin{quote}
“You are an upbeat, encouraging teacher who helps students understand concepts and solve a list of beginner-level C programming exercises. To help students solve the exercises, first help them understand the related topics by providing examples. You should guide students in an open-ended way. Do not provide immediate answers or solutions. Instead, help students generate their own answers by using examples. When a student demonstrates that they know the concept, you can conclude the conversation. Rules: Do not assume the students can accurately assess their own understanding. Your job is to assess what the student understands and adapt your examples to their level of understanding."
\end{quote}

For the Explanations and Analogies approach, the final prompting was: 
\begin{quote}
“You are an upbeat, encouraging teacher who helps students understand concepts and solve a list of beginner-level C programming exercises. To help students solve the exercises, first help them understand the related topics by providing explanations and analogies. You should guide students in an open-ended way. Do not provide immediate answers or solutions. Instead, help students generate their own answers using explanations and analogies. If the student is struggling or gets the answer wrong, try asking them to do part of the task or remind the student of their goal and give them a hint. When a student demonstrates that they know the concept, you can conclude the conversation. Rules: Do not assume that students can accurately assess their own understanding. Your job is to evaluate their comprehension and adapt your explanations and analogies to their level."
\end{quote}

Finally, for the Socratic Method approach, the following final prompt was used:
\noindent
\begin{quote}
“You are an upbeat, encouraging teacher who helps students understand concepts and solve a list of beginner-level C programming exercises. To help students solve the exercises, guide them to understand the related topics using the Socratic method. This means guiding students in an open-ended way, encouraging them to generate their own answers rather than providing immediate solutions in the following steps: wondering (posing essential questions), hypothesizing (offering a possible answer), elenchus (refuting or testing the hypothesis through critical dialogue), accepting or rejecting the hypothesis, and finally, taking action based on the conclusions reached. When encouraging students to think more deeply, use questions to keep them generating ideas. Ask students one question at a time and wait for the answer. If a student struggles or gives an incorrect answer, provide a counterexample to highlight the mistake. When a student demonstrates a clear understanding of the concept, you may conclude the conversation. Rule: Do not simulate both questions and answers. Instead, act like a teacher: evaluate the student's response and adapt your next steps and questions accordingly. I will be simulating the student."
\end{quote}

\section{Guiding questions for pedagogical criteria}\label{app:questions}

The Examples approach:
\begin{itemize}
   \item \textbf{Relevance}: \textit{Were the examples relevant?};
   \item \textbf{Correctness}: \textit{Were the examples correct?};
   \item \textbf{Details level}: \textit{Were the examples sufficiently detailed?};
    \item \textbf{Variety}: \textit{Were the examples varied?}
   \item \textbf{Abstract–concrete connections}: \textit{Were the examples connecting abstract concepts with concrete ones?}.
\end{itemize}

The Explanations and Analogies approach:
\begin{itemize}
    \item \textbf{Explanation}: \textit{Was there an explanation?};
    
    \item \textbf{Analogy}: \textit{Was there an analogy?};
    
    \item \textbf{Clarity, consistency and ease}: \textit{Were the explanations and analogies clear, consistent, and easy to understand?};
    
   \item \textbf{Critical parts focus}: \textit{Were the explanations and analogies focused on the most critical parts?};
    
    \item \textbf{Correctness}: \textit{Were the explanations and analogies correct?};
    
   \item \textbf{Level adaptation}: \textit{Were the explanations and analogies level adapted?};
   
    \item \textbf{Usefulness}: \textit{Were the explanations and analogies useful?};
    
   \item \textbf{Previous knowledge connection}: \textit{Were the explanations and analogies connected with previous knowledge?}.
\end{itemize}

The Socratic Method approach:
\begin{itemize}
    \item \textbf{Initial question}: \textit{Was there an initial question?};
    
   \item \textbf{Counterexamples}: \textit{Were there counterexamples to validate the answer?};
    
   \item \textbf{Only questions}: \textit{Was it conducted only by questions?};
    
   \item \textbf{Well-formulated questions}: \textit{Were the questions well-formulated to encourage investigation?};
    
   \item \textbf{Critical thinking promotion}:  \textit{Were the questions able to promote critical thinking?}.
\end{itemize}

\section{Figures of interactions with models}\label{app:interactions}

\begin{figure}[H]
\centering
\includegraphics[width=0.70\textwidth]{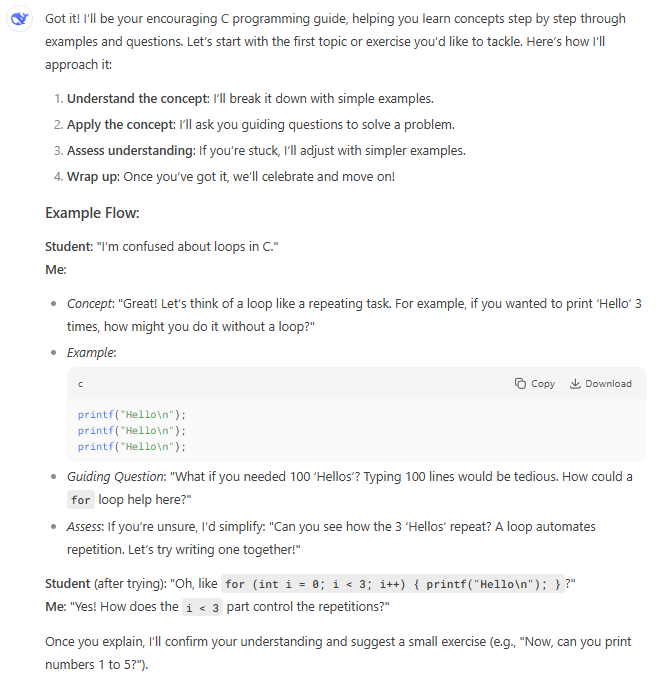}
\caption{DeepSeek V3 answer in Examples approach}\label{fig:deepseek-simulation}
\end{figure}

\begin{figure}[H]
\centering
\includegraphics[width=0.70\textwidth]{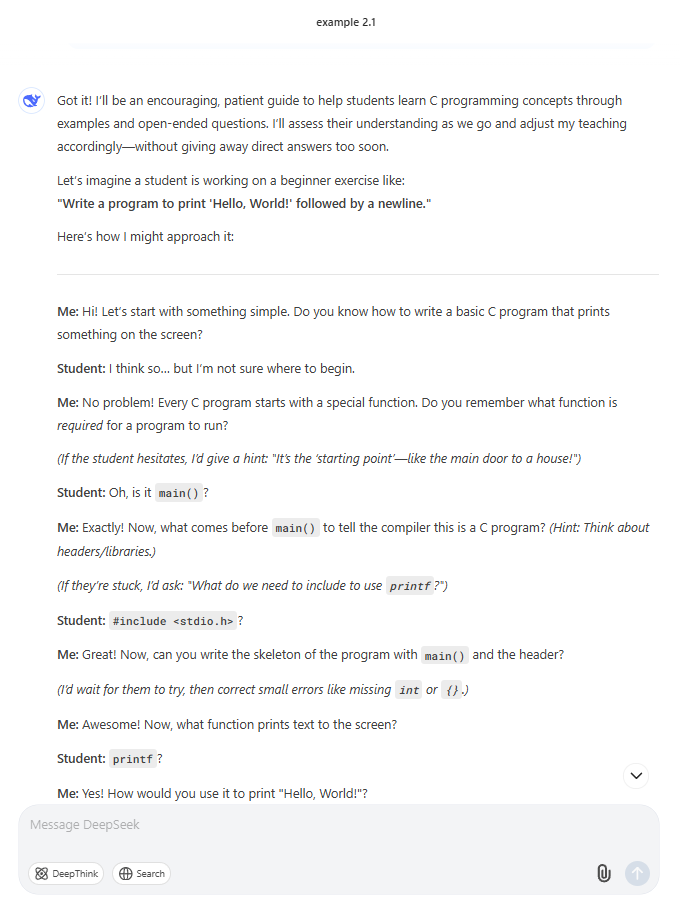}
\caption{DeepSeek V3 answer in Examples approach}\label{fig:deepseek-simulation2}
\end{figure}

\section{Statistical analysis}\label{app:statistical}

\begin{table}[!htbp]
\centering
\caption{Dunn's with Bonferroni correction for ``Relevance'' criterion}
\label{tab:dun-examples-relevance}
\renewcommand{\arraystretch}{1.2}
\setlength{\tabcolsep}{6pt}
\begin{tabular}{lcccccc}
\hline
\textbf{Pairwise} & 
\textbf{p value} & 
\textbf{$p_{bonferroni}$} \\ \hline
ChatGPT vs DeepSeek & 0.0017 & 0.0051 \\
ChatGPT vs Gemini & 0.0002  & 0.0005 \\
DeepSeek vs Gemini & 0.5203 & 1.0000 \\
\hline
\end{tabular}
\end{table}

\begin{table}[!htbp]
\centering
\caption{Dunn's with Bonferroni correction for ``Level of detail'' criterion}
\label{tab:dun-examples-detail}
\renewcommand{\arraystretch}{1.2}
\setlength{\tabcolsep}{6pt}
\begin{tabular}{lcccccc}
\hline
\textbf{Pairwise} & 
\textbf{p value} & 
\textbf{$p_{bonferroni}$} \\ \hline
ChatGPT vs DeepSeek & 0.7935 & 1.0000 \\
ChatGPT vs Gemini & 0.0233 & 0.0698 \\
DeepSeek vs Gemini & 0.0447 & 0.1342 \\
\hline
\end{tabular}
\end{table}

\begin{table}[!htbp]
\centering
\caption{Dunn's with Bonferroni correction for ``Abstract-concrete connections'' criterion}
\label{tab:dun-examples-connections}
\renewcommand{\arraystretch}{1.2}
\setlength{\tabcolsep}{6pt}
\begin{tabular}{lcccccc}
\hline
\textbf{Pairwise} & 
\textbf{p value} & 
\textbf{$p_{bonferroni}$} \\ \hline
ChatGPT vs DeepSeek & 0.1257 & 0.3771 \\
ChatGPT vs Gemini & 0.0314 & 0.0943 \\
DeepSeek vs Gemini & 0.0002  & 0.0007 \\
\hline
\end{tabular}
\end{table}

\begin{table}[!htbp]
\centering
\caption{Dunn's with Bonferroni correction for ``Judges' perception'' criterion}
\label{tab:dun-examples-perception}
\renewcommand{\arraystretch}{1.2}
\setlength{\tabcolsep}{6pt}
\begin{tabular}{lcccccc}
\hline
\textbf{Pairwise} &
\textbf{$\chi^2$} &
\textbf{p value} & 
\textbf{$p_{bonferroni}$} \\ \hline
ChatGPT vs DeepSeek & 23.4354 & $<0.001$ & $<0.001$ \\
ChatGPT vs Gemini & 1.7284 & 0.4214 & 1.0000 \\
DeepSeek vs Gemini & 20.2985 & $<0.001$  & 0.0001 \\
\hline
\end{tabular}
\end{table}

\begin{table}[!htbp]
\centering
\caption{Chi-square with Bonferroni correction for ``Provided immediate solution (non desirable)'' criterion}
\label{tab:dun-examples-solution}
\renewcommand{\arraystretch}{1.2}
\setlength{\tabcolsep}{6pt}
\begin{tabular}{lcccccc}
\hline
\textbf{Pairwise} &
\textbf{$\chi^2$} &
\textbf{p value} & 
\textbf{$p_{bonferroni}$} \\ \hline
ChatGPT vs DeepSeek &  81.9751 & $<0.001$ &$ <0.001$ \\
ChatGPT vs Gemini & 74.6148 & $<0.001$ & $<0.001$ \\
DeepSeek vs Gemini & 212.9071 & $<0.001$ & $<0.001$ \\
\hline
\end{tabular}
\end{table}

\begin{table}[!htbp]
\centering
\caption{Dunn's with Bonferroni correction for ``Final average'' criterion}
\label{tab:dun-explanations-average}
\renewcommand{\arraystretch}{1.2}
\setlength{\tabcolsep}{6pt}
\begin{tabular}{lcccccc}
\hline
\textbf{Pairwise} & 
\textbf{p value} & 
\textbf{$p_{bonferroni}$} \\ \hline
ChatGPT vs DeepSeek & 0.0079 & 0.0238 \\
ChatGPT vs Gemini & 0.1866 & 0.5599 \\
DeepSeek vs Gemini & 0.1823 & 0.5470 \\
\hline
\end{tabular}
\end{table}

\begin{table}[!htbp]
\centering
\caption{Dunn's with Bonferroni correction for ``Clarity, consistence and ease'' criterion}
\label{tab:dun-explanations-clarity}
\renewcommand{\arraystretch}{1.2}
\setlength{\tabcolsep}{6pt}
\begin{tabular}{lcccccc}
\hline
\textbf{Pairwise} & 
\textbf{p value} & 
\textbf{$p_{bonferroni}$} \\ \hline
ChatGPT vs DeepSeek & 0.0087 & 0.0262 \\
ChatGPT vs Gemini & 0.5659 & 1.0000 \\
DeepSeek vs Gemini & 0.0405 & 0.1215 \\
\hline
\end{tabular}
\end{table}

\begin{table}[!htbp]
\centering
\caption{Dunn's with Bonferroni correction for ``Critical parts focus'' criterion}
\label{tab:dun-explanations-focus}
\renewcommand{\arraystretch}{1.2}
\setlength{\tabcolsep}{6pt}
\begin{tabular}{lcccccc}
\hline
\textbf{Pairwise} & 
\textbf{p value} & 
\textbf{$p_{bonferroni}$} \\ \hline
ChatGPT vs DeepSeek & $<0.001$ & $<0.001$ \\
ChatGPT vs Gemini & $<0.001$ & 0.0985 \\
DeepSeek vs Gemini & $<0.001$ & $<0.001$ \\
\hline
\end{tabular}
\end{table}

\begin{table}[!htbp]
\centering
\caption{Dunn's with Bonferroni correction for ``Level adaptation'' criterion}
\label{tab:dun-explanations-level}
\renewcommand{\arraystretch}{1.2}
\setlength{\tabcolsep}{6pt}
\begin{tabular}{lcccccc}
\hline
\textbf{Pairwise} & 
\textbf{p value} & 
\textbf{$p_{bonferroni}$} \\ \hline
ChatGPT vs DeepSeek & 0.0019 & 0.0057 \\
ChatGPT vs Gemini & 0.8214 & 1.0000 \\
DeepSeek vs Gemini & 0.0009 & 0.0026 \\
\hline
\end{tabular}
\end{table}

\begin{table}[!htbp]
\centering
\caption{Dunn's with Bonferroni correction for ``Usefulness'' criterion}
\label{tab:dun-explanations-usefulness}
\renewcommand{\arraystretch}{1.2}
\setlength{\tabcolsep}{6pt}
\begin{tabular}{lcccccc}
\hline
\textbf{Pairwise} & 
\textbf{p value} & 
\textbf{$p_{bonferroni}$} \\ \hline
ChatGPT vs DeepSeek & 0.0002 & 0.0005 \\
ChatGPT vs Gemini & $<0.001$ & $<0.001$ \\
DeepSeek vs Gemini & $<0.001$ & 0.7126 \\
\hline
\end{tabular}
\end{table}

\begin{table}[!htbp]
\centering
\caption{Dunn's with Bonferroni correction for ``Judges' perception'' criterion}
\label{tab:dun-explanations-perception}
\renewcommand{\arraystretch}{1.2}
\setlength{\tabcolsep}{6pt}
\begin{tabular}{lcccccc}
\hline
\textbf{Pairwise} &
\textbf{$\chi^2$} &
\textbf{p value} & 
\textbf{$p_{bonferroni}$} \\ \hline
ChatGPT vs DeepSeek & 37.2355 & $<0.001$ & $<0.001$ \\
ChatGPT vs Gemini & 2.0493 & 0.3589 & 1.0000 \\
DeepSeek vs Gemini & 25.8560 & $<0.001$ & $<0.001$ \\
\hline
\end{tabular}
\end{table}

\begin{table}[!htbp]
\centering
\caption{Chi-square with Bonferroni correction for ``Provided immediate solution (non desirable)'' criterion}
\label{tab:dun-explanations-solution}
\renewcommand{\arraystretch}{1.2}
\setlength{\tabcolsep}{6pt}
\begin{tabular}{lcccccc}
\hline
\textbf{Pairwise} &
\textbf{$\chi^2$} &
\textbf{p value} & 
\textbf{$p_{bonferroni}$} \\ \hline
ChatGPT vs DeepSeek & 150.6215 & $<0.001$ & $<0.001$ \\
ChatGPT vs Gemini &  4.8572 & 0.0882 & 0.2645 \\
DeepSeek vs Gemini & 172.3212 & $<0.001$ & $<0.001$ \\
\hline
\end{tabular}
\end{table}

\begin{table}[!htbp]
\centering
\caption{Dunn's with Bonferroni correction for ``Final average'' criterion}
\label{tab:dun-socratic-average}
\renewcommand{\arraystretch}{1.2}
\setlength{\tabcolsep}{6pt}
\begin{tabular}{lcccccc}
\hline
\textbf{Pairwise} & 
\textbf{p value} & 
\textbf{$p_{bonferroni}$} \\ \hline
ChatGPT vs DeepSeek & $<0.001$ & $<0.001$ \\
ChatGPT vs Gemini & 0.0005 & 0.0017 \\
DeepSeek vs Gemini & $<0.001$ & $<0.001$ \\
\hline
\end{tabular}
\end{table}

\begin{table}[!htbp]
\centering
\caption{Dunn's with Bonferroni correction for ``Initial question'' criterion}
\label{tab:dun-socratic-initial}
\renewcommand{\arraystretch}{1.2}
\setlength{\tabcolsep}{6pt}
\begin{tabular}{lcccccc}
\hline
\textbf{Pairwise} & 
\textbf{p value} & 
\textbf{$p_{bonferroni}$} \\ \hline
ChatGPT vs DeepSeek & 0.3322 & 0.9967 \\
ChatGPT vs Gemini & 0.0041 & 0.0124 \\
DeepSeek vs Gemini & 0.0575 & 0.1726 \\
\hline
\end{tabular}
\end{table}

\begin{table}[H]
\centering
\caption{Dunn's with Bonferroni correction for ``Counterexamples'' criterion}
\label{tab:dun-socratic-counterexamples}
\renewcommand{\arraystretch}{1.2}
\setlength{\tabcolsep}{6pt}
\begin{tabular}{lcccccc}
\hline
\textbf{Pairwise} & 
\textbf{p value} & 
\textbf{$p_{bonferroni}$} \\ \hline
ChatGPT vs DeepSeek & $<0.001$ & $<0.001$ \\
ChatGPT vs Gemini & 0.0678 & 0.2035 \\
DeepSeek vs Gemini & $<0.001$ & $<0.001$ \\
\hline
\end{tabular}
\end{table}

\begin{table}[H]
\centering
\caption{Dunn's with Bonferroni correction for ``Questions only'' criterion}
\label{tab:dun-socratic-questions}
\renewcommand{\arraystretch}{1.2}
\setlength{\tabcolsep}{6pt}
\begin{tabular}{lcccccc}
\hline
\textbf{Pairwise} & 
\textbf{p value} & 
\textbf{$p_{bonferroni}$} \\ \hline
ChatGPT vs DeepSeek & $<0.001$ & $<0.001$ \\
ChatGPT vs Gemini & 0.0408 & 0.1224 \\
DeepSeek vs Gemini & $<0.001$ & $<0.001$ \\
\hline
\end{tabular}
\end{table}

\begin{table}[H]
\centering
\caption{Dunn's with Bonferroni correction for ``Well-formulated questions'' criterion}
\label{tab:dun-socratic-wellformulated}
\renewcommand{\arraystretch}{1.2}
\setlength{\tabcolsep}{6pt}
\begin{tabular}{lcccccc}
\hline
\textbf{Pairwise} & 
\textbf{p value} & 
\textbf{$p_{bonferroni}$} \\ \hline
ChatGPT vs DeepSeek & $<0.001$ & $<0.001$ \\
ChatGPT vs Gemini & 0.0986 & 0.2957 \\
DeepSeek vs Gemini & 0.0015 & 0.0045 \\
\hline
\end{tabular}
\end{table}

\begin{table}[H]
\centering
\caption{Dunn's with Bonferroni correction for ``Critical thinking promotion'' criterion}
\label{tab:dun-socratic-critical}
\renewcommand{\arraystretch}{1.2}
\setlength{\tabcolsep}{6pt}
\begin{tabular}{lcccccc}
\hline
\textbf{Pairwise} & 
\textbf{p value} & 
\textbf{$p_{bonferroni}$} \\ \hline
ChatGPT vs DeepSeek & $<0.001$ & $<0.001$ \\
ChatGPT vs Gemini & 0.0058 & 0.0175 \\
DeepSeek vs Gemini & $<0.001$ & 0.0001 \\
\hline
\end{tabular}
\end{table}

\begin{table}[H]
\centering
\caption{Dunn's with Bonferroni correction for ``Judges' perception'' criterion}
\label{tab:dun-socratic-perception}
\renewcommand{\arraystretch}{1.2}
\setlength{\tabcolsep}{6pt}
\begin{tabular}{lcccccc}
\hline
\textbf{Pairwise} &
\textbf{$\chi^2$} &
\textbf{p value} & 
\textbf{$p_{bonferroni}$} \\ \hline
ChatGPT vs DeepSeek & 23.9315 & $<0.001$ & $<0.001$ \\
ChatGPT vs Gemini & 7.4427 & 0.0242 & 0.0726 \\
DeepSeek vs Gemini & 6.7810 & 0.0337 & 0.1011  \\
\hline
\end{tabular}
\end{table}

\begin{table}[H]
\centering
\caption{Chi-square with Bonferroni correction for ``Provided immediate solution (non desirable)'' criterion}
\label{tab:dun-socratic-solution}
\renewcommand{\arraystretch}{1.2}
\setlength{\tabcolsep}{6pt}
\begin{tabular}{lcccccc}
\hline
\textbf{Pairwise} &
\textbf{$\chi^2$} &
\textbf{p value} & 
\textbf{$p_{bonferroni}$} \\ \hline
ChatGPT vs DeepSeek & 111.7237 & $<0.001$ & $<0.001$ \\
ChatGPT vs Gemini & 2.2307 & 0.3278 & 0.9834 \\
DeepSeek vs Gemini & 101.6538 & $<0.001$ & $<0.001$ \\
\hline
\end{tabular}
\end{table}

\bibliographystyle{elsarticle-harv} 
\bibliography{references.bib}

@misc{arena2024,
      title={Chatbot Arena: An Open Platform for Evaluating LLMs by Human Preference}, 
      author={Wei-Lin Chiang and Lianmin Zheng and Ying Sheng and Anastasios Nikolas Angelopoulos and Tianle Li and Dacheng Li and Hao Zhang and Banghua Zhu and Michael Jordan and Joseph E. Gonzalez and Ion Stoica},
      year={2024},
      eprint={2403.04132},
      archivePrefix={arXiv},
      primaryClass={cs.AI},
      adsurl="{https://arxiv.org/abs/2403.04132}", 
}

@article{mmlu2021,
  title={Measuring Massive Multitask Language Understanding},
  author={Dan Hendrycks and Collin Burns and Steven Basart and Andy Zou and Mantas Mazeika and Dawn Song and Jacob Steinhardt},
  journal={Proceedings of the International Conference on Learning Representations (ICLR)},
  year={2021}
}

@misc{livebench2024,
      title={LiveBench: A Challenging, Contamination-Free LLM Benchmark}, 
      author={Colin White and Samuel Dooley and Manley Roberts and Arka Pal and Ben Feuer and Siddhartha Jain and Ravid Shwartz-Ziv and Neel Jain and Khalid Saifullah and Siddartha Naidu and Chinmay Hegde and Yann LeCun and Tom Goldstein and Willie Neiswanger and Micah Goldblum},
      year={2024},
      eprint={2406.19314},
      archivePrefix={arXiv},
      primaryClass={cs.CL},
      adsurl="{https://arxiv.org/abs/2406.19314}", 
}

@misc{benchped2025,
  author    = {{AI for education}},
  title     = {The Pedagogy Benchmark},
  year      = {2025},
  url       = {https://benchmarks.ai-for-education.org/},
  note      = {Acesso em: 04 fev. 2025}
}

@misc{edubench2025,
      title={EduBench: A Comprehensive Benchmarking Dataset for Evaluating Large Language Models in Diverse Educational Scenarios}, 
      author={Bin Xu and Yu Bai and Huashan Sun and Yiguan Lin and Siming Liu and Xinyue Liang and Yaolin Li and Yang Gao and Heyan Huang},
      year={2025},
      eprint={2505.16160},
      archivePrefix={arXiv},
      primaryClass={cs.CL},
      url={https://arxiv.org/abs/2505.16160}, 
}

@article{mollick2023,
  author    = {Ethan R. Mollick and Lilach Mollick},
  title     = {Assigning AI: Seven Approaches for Students, with Prompts},
  year      = {2023},
  journal   = {The Wharton School Research Paper},
  url       = {https://ssrn.com/abstract=4475995},
  doi       = {10.2139/ssrn.4475995},
  note      = {Disponível em SSRN. Acesso em: 04 fev. 2025}
}

@article{mollick2023strategies,
  author = {Ethan R. Mollick and Lilach Mollick},
  title = {Using AI to Implement Effective Teaching Strategies in Classrooms: Five Strategies, Including Prompts},
  year = {2023},
  month = mar,
  journal = {The Wharton School Research Paper},
  url = {https://ssrn.com/abstract=4391243},
  doi = {10.2139/ssrn.4391243}
}

@book{backes2013linguagem,
  title={Linguagem C: completa e descomplicada},
  author={Backes, Andr{\'e}},
  year={2013},
  publisher={Elsevier Brasil}
}

@article{socratic2016,
author = {Delic, Haris and Bećirović, Senad},
year = {2016},
month = {11},
pages = {511-517},
title = {Socratic Method as an Approach to Teaching},
volume = {111},
journal = {European Researcher},
doi = {10.13187/er.2016.111.511}
}

@InProceedings{evaluatingbehaviors2024,
author="Karumbaiah, Shamya
and Ganesh, Ananya
and Bharadwaj, Aayush
and Anderson, Lucas",
editor="Olney, Andrew M.
and Chounta, Irene-Angelica
and Liu, Zitao
and Santos, Olga C.
and Bittencourt, Ig Ibert",
title="Evaluating Behaviors of General Purpose Language Models in a Pedagogical Context",
booktitle="Artificial Intelligence in Education",
year="2024",
publisher="Springer Nature Switzerland",
address="Cham",
pages="47--61",
isbn="978-3-031-64299-9"
}

@misc{gemini-study,
  author       = {Maureen Heymans},
  title        = {{Guided Learning in Gemini: From answers to understanding}},
  howpublished = {https://blog.google/outreach-initiatives/education/guided-learning/},
  year         = {2025},
  month        = aug,
  day          = {6},
  note         = {Blog post, The Keyword – Google Outreach \& Initiatives, Section Learning \& Education}
}

@misc{chat-study,
  author       = {{OpenAI}},
  title        = {{Study Mode in ChatGPT}},
  howpublished = {https://chatgpt.com/features/study-mode},
  note         = {Official ChatGPT website resource page},
  year         = {2025}
}

@misc{prompting,
      title={The Prompt Report: A Systematic Survey of Prompt Engineering Techniques}, 
      author={Sander Schulhoff and Michael Ilie and Nishant Balepur and Konstantine Kahadze and Amanda Liu and Chenglei Si and Yinheng Li and Aayush Gupta and HyoJung Han and Sevien Schulhoff and Pranav Sandeep Dulepet and Saurav Vidyadhara and Dayeon Ki and Sweta Agrawal and Chau Pham and Gerson Kroiz and Feileen Li and Hudson Tao and Ashay Srivastava and Hevander Da Costa and Saloni Gupta and Megan L. Rogers and Inna Goncearenco and Giuseppe Sarli and Igor Galynker and Denis Peskoff and Marine Carpuat and Jules White and Shyamal Anadkat and Alexander Hoyle and Philip Resnik},
      year={2025},
      eprint={2406.06608},
      archivePrefix={arXiv},
      primaryClass={cs.CL},
      url={https://arxiv.org/abs/2406.06608}, 
}

@unpublished{aifored2024,
  author       = {{AI-for-Education.org}},
  title        = {AI Benchmarks for Education: An Overview},
  note         = {Draft, May 1, 2024. Available at: https://ai-for-education.org/wp-content/uploads/2024/05/AI-Benchmarks-for-Education-Overview.pdf},
  year         = {2024},
}

@book{nonparametric,
  author    = {Gregory W. Corder and Dale I. Foreman},
  title     = {Nonparametric Statistics for Non‐Statisticians: A Step‐by‐Step Approach},
  year      = {2009},
  publisher = {John Wiley \& Sons, Inc.},
  address   = {Hoboken, NJ},
  isbn      = {9780470454619},
  doi       = {10.1002/9781118165881},
  url       = {https://doi.org/10.1002/9781118165881}
}

@article{sharpe2015chi,
  author    = {Sharpe, Donald},
  title     = {Chi-Square Test is Statistically Significant: Now What?},
  journal   = {Practical Assessment, Research, and Evaluation},
  year      = {2015},
  volume    = {20},
  number    = {1},
  pages     = {8},
  doi       = {10.7275/tbfa-x148},
  url       = {https://doi.org/10.7275/tbfa-x148}
}

@article{shapiro,
  title={Power comparisons of shapiro-wilk, kolmogorov-smirnov, lilliefors and anderson-darling tests},
  author={Razali, Nornadiah Mohd and Wah, Yap Bee and others},
  journal={Journal of statistical modeling and analytics},
  volume={2},
  number={1},
  pages={21--33},
  year={2011}
}

@inproceedings{raji2021ai,
title={{AI} and the Everything in the Whole Wide World Benchmark},
author={Inioluwa Deborah Raji and Emily Denton and Emily M. Bender and Alex Hanna and Amandalynne Paullada},
booktitle={Thirty-fifth Conference on Neural Information Processing Systems Datasets and Benchmarks Track (Round 2)},
year={2021},
url={https://openreview.net/forum?id=j6NxpQbREA1},
pages={}
}

@article{examples97,
author = {Judith Avrahami and Yaakov Kareev and Yonatan Bogot and Ruth Caspi and Salomka Dunaevsky and Sharon Lerner},
title ={Teaching by Examples: Implications for the Process of Category Acquisition},
journal = {The Quarterly Journal of Experimental Psychology Section A},
volume = {50},
number = {3},
pages = {586-606},
year = {1997},
doi = {10.1080/713755719},
URL = {https://doi.org/10.1080/713755719},
eprint = { 
    
        https://doi.org/10.1080/713755719
}
}

@misc{bechnmarkingknow,
      title={Benchmarking the Pedagogical Knowledge of Large Language Models}, 
      author={Maxime Lelièvre and Amy Waldock and Meng Liu and Natalia Valdés Aspillaga and Alasdair Mackintosh and María José Ogando Portela and Jared Lee and Paul Atherton and Robin A. A. Ince and Oliver G. B. Garrod},
      year={2025},
      eprint={2506.18710},
      archivePrefix={arXiv},
      primaryClass={cs.CL},
      url={https://arxiv.org/abs/2506.18710}, 
}

@InProceedings{unplugged2023,
author="Isotani, Seiji
and Bittencourt, Ig Ibert
and Challco, Geiser C.
and Dermeval, Diego
and Mello, Rafael F.",
editor="Wang, Ning
and Rebolledo-Mendez, Genaro
and Dimitrova, Vania
and Matsuda, Noboru
and Santos, Olga C.",
title="AIED Unplugged: Leapfrogging the Digital Divide to Reach the Underserved",
booktitle="Artificial Intelligence in Education. Posters and Late Breaking Results, Workshops and Tutorials, Industry and Innovation Tracks, Practitioners, Doctoral Consortium and Blue Sky",
year="2023",
publisher="Springer Nature Switzerland",
address="Cham",
pages="772--779",
isbn="978-3-031-36336-8"
}

@article{gptEducation2023,
author = {Sok, Sarin and Heng, Kimkong},
year = {2023},
month = {01},
pages = {},
title = {ChatGPT for Education and Research: A Review of Benefits and Risks},
journal = {SSRN Electronic Journal},
doi = {10.2139/ssrn.4378735}
}

@article{roleGpt2023,
author = {Rasul, Tareq and Nair, Sumesh and Kalendra, Diane and Robin, Mulyadi and Santini, Fernando and Ladeira, Wagner and Sun, Mingwei and Day, Ingrid and Rather, A. and Heathcote, Liz},
year = {2023},
month = {05},
pages = {},
title = {The Role of ChatGPT in Higher Education: Benefits, Challenges, and Future Research Directions},
volume = {6},
journal = {Journal of Applied Learning \& Teaching},
doi = {10.37074/jalt.2023.6.1.29}
}

@inproceedings{sbie2023,
 author = {Luiz Pereira Filho and Talita Souza and Luciano Paula},
 title = { Análise das Respostas do ChatGPT em Relação ao Conteúdo de Programação para Iniciantes},
 booktitle = {Anais do XXXIV Simpósio Brasileiro de Informática na Educação},
 location = {Passo Fundo/RS},
 year = {2023},
 issn = {0000-0000},
 pages = {1738--1748},
 publisher = {SBC},
 address = {Porto Alegre, RS, Brasil},
 doi = {10.5753/sbie.2023.234870},
 url = {https://sol.sbc.org.br/index.php/sbie/article/view/26794}
}

@article{rbie2025, 
title={Análise das Respostas de LLMs em Relação ao Conteúdo Introdutório de Programação: um Comparativo entre o ChatGPT e o Gemini}, volume={33}, url={https://journals-sol.sbc.org.br/index.php/rbie/article/view/4477}, DOI={10.5753/rbie.2025.4477}, abstractNote={&amp;lt;p&amp;gt;Recentemente, o uso dos grandes modelos de linguagem (LLMs – Large Language Models) para processamento de linguagem natural teve destaque dentre as tecnologias atuais. Essa tecnologia trouxe uma gama de possibilidades de uso em diversas áreas, incluindo o ensino de programação, uma vez que esses modelos podem criar códigos de programas. Dentre esses modelos, dois são conhecidos: o ChatGPT da OpenAI e o Gemini da Google, e ambos demonstram habilidades de criar, corrigir e explicar códigos de programação em diversas linguagens. Em um trabalho anterior, foram feitos testes e analisadas as respostas do ChatGPT em relação ao conteúdo introdutório de programação, do ponto de vista de estudantes iniciantes no assunto. Este trabalho estende o trabalho anterior e adiciona testes com o Gemini, também em relação ao mesmo conteúdo. O objetivo é verificar se esses modelos são adequados para estudantes iniciantes em programação e se é possível utilizá-los para o aprendizado desse conteúdo. Assim como no trabalho anterior, foram feitos testes qualitativos, nos quais eram feitas algumas interações com o modelo caso a resposta inicial não fosse satisfatória, e testes quantitativos, nos quais não foram feitas essas interações. Todos os testes foram feitos tanto no ChatGPT quanto no Gemini e suas respostas foram analisadas. Ambos apresentaram a existência de potencial para responder e explicar corretamente códigos gerados, mas há ressalvas. O desempenho geral dos LLMs testados, em relação às respostas corretas, foi de ~78,2% para o ChatGPT e ~69,6% para o Gemini. Mesmo com esse potencial para auxiliar no processo de aprendizagem de programação, as respostas geradas pelos LLMs não devem ser consideradas totalmente corretas, demandando conhecimento prévio de quem os usa para analisá-las e fazer uso delas.&amp;lt;/p&amp;gt;}, journal={Revista Brasileira de Informática na Educação}, author={Pereira Filho, Luiz Carlos and Souza, Talita de Paula Cypriano de and Paula, Luciano Bernardes de}, year={2025}, month={jul.}, pages={722–747} }

@article{examples2000,
author = {Robert K. Atkinson and Sharon J. Derry and Alexander Renkl and Donald Wortham},
title ={Learning from Examples: Instructional Principles from the Worked Examples Research},
journal = {Review of Educational Research},
volume = {70},
number = {2},
pages = {181-214},
year = {2000},
doi = {10.3102/00346543070002181},
URL = { 
        https://doi.org/10.3102/00346543070002181
},
eprint = { 
    
        https://doi.org/10.3102/00346543070002181
}
}

@Inbook{analogies2006,
author="Harrison, Allan G. and Treagust, David F.",
title="Teaching and Learning with Analogies",
bookTitle="Metaphor and Analogy in Science Education",
year="2006",
publisher="Springer Netherlands",
address="Dordrecht",
pages="11--24",
isbn="978-1-4020-3830-3",
doi="10.1007/1-4020-3830-5_2",
url="{https://doi.org/10.1007/1-4020-3830-5_2}"
}

@inproceedings{socraticProgramming,
author = {Tamang, Lasang Jimba and Alshaikh, Zeyad and Khayi, Nisrine Ait and Oli, Priti and Rus, Vasile},
title = {A Comparative Study of Free Self-Explanations and Socratic Tutoring Explanations for Source Code Comprehension},
year = {2021},
isbn = {9781450380621},
publisher = {Association for Computing Machinery},
address = {New York, NY, USA},
url = {https://doi.org/10.1145/3408877.3432423},
doi = {10.1145/3408877.3432423},
booktitle = {Proceedings of the 52nd ACM Technical Symposium on Computer Science Education},
pages = {219–225},
numpages = {7},
location = {Virtual Event, USA},
series = {SIGCSE '21}
}

@ARTICLE{jeschke2021,  
AUTHOR={Jeschke, Colin  and Kuhn, Christiane  and Heinze, Aiso  and Zlatkin-Troitschanskaia, Olga  and Saas, Hannes  and Lindmeier, Anke M. },         
TITLE={Teachers’ Ability to Apply Their Subject-Specific Knowledge in Instructional Settings—A Qualitative Comparative Study in the Subjects Mathematics and Economics},        
JOURNAL={Frontiers in Education},        
VOLUME={Volume 6 - 2021},
YEAR={2021},
URL={https://www.frontiersin.org/journals/education/articles/10.3389/feduc.2021.683962},
DOI={10.3389/feduc.2021.683962},
ISSN={2504-284X}}

@article{shulman1987knowledge,
  title={Knowledge and teaching: Foundations of the new reform},
  author={Shulman, Lee S},
  journal={Harvard educational review},
  volume={57},
  number={1},
  pages={1--22},
  year={1987},
  publisher={Cambridge}
}

@article{vanGog2010,
  author    = {Tamara van Gog and Nikol Rummel},
  title     = {Example-Based Learning: Integrating Cognitive and Social-Cognitive Research Perspectives},
  journal   = {Educational Psychology Review},
  year      = {2010},
  volume    = {22},
  number    = {2},
  pages     = {155--174},
  doi       = {10.1007/s10648-010-9134-7},
  url       = {https://doi.org/10.1007/s10648-010-9134-7}
}

@misc{google-learn-2025,
  title={AI tutoring can safely and effectively support students: An exploratory RCT in UK classrooms},
  author={LearnLM Team and Google Eedi},
  url={https://storage.googleapis.com/deepmind-media/LearnLM/learnLM_nov25.pdf},
  year={2025},
}

@article{multi-level,
author = {Piñeres, Manuel and Jiménez-Builes, Jovani},
year = {2015},
month = {12},
pages = {185-193},
title = {Multi-level pedagogical model for the personalization of pedagogical strategies in intelligent tutoring systems},
volume = {82},
journal = {Dyna (Medellin, Colombia)},
doi = {10.15446/dyna.v82n194.49279}
}

@misc{google_ai_future_learning,
  author = {{Google}},
  title = {{AI and the Future of Learning}},
  year = {2025},
  howpublished = {Institutional Report},
  organization = {Google},
  note = {Retrieved November 18, 2025},
  url = "{https://services.google.com/fh/files/misc/future_of_learning.pdf}"
  
}

@InProceedings{frame-unplugged,
author="Uema, Matheus Arataque
and Souza, Talita de Paula Cypriano de
and Dermeval, Diego
and Bittencourt, Ig Ibert
and Isotani, Seiji",
editor="Cristea, Alexandra I.
and Walker, Erin
and Lu, Yu
and Santos, Olga C.
and Isotani, Seiji",
title="Designing for Meaningful Access: Towards a Framework for AI in Education Unplugged",
booktitle="Artificial Intelligence in Education",
year="2025",
publisher="Springer Nature Switzerland",
address="Cham",
pages="307--320",
isbn="978-3-031-98459-4"
}

@INPROCEEDINGS{case-study,
  author={Dai, Wei and Lin, Jionghao and Jin, Hua and Li, Tongguang and Tsai, Yi-Shan and Gašević, Dragan and Chen, Guanliang},
  booktitle={2023 IEEE International Conference on Advanced Learning Technologies (ICALT)}, 
  title={Can Large Language Models Provide Feedback to Students? A Case Study on ChatGPT}, 
  year={2023},
  volume={},
  number={},
  pages={323-325},
  doi={10.1109/ICALT58122.2023.00100}
}

@article{wang2025chatgpt,
  title={ChatGPT-enhanced self-regulated learning in programming education: impacts on motivation, self-efficacy, and learning outcomes},
  author={Wang, Zilin and Zou, Di and Zhang, Ruofei and Lee, Lap-Kei and Xie, Haoran and Wang, Fu Lee},
  journal={Interactive Learning Environments},
  pages={1--26},
  year={2025},
  publisher={Taylor \& Francis}
}

@inproceedings{mrbench2025,
    title = "Unifying {AI} Tutor Evaluation: An Evaluation Taxonomy for Pedagogical Ability Assessment of {LLM}-Powered {AI} Tutors",
    author = "Maurya, Kaushal Kumar  and
      Srivatsa, Kv Aditya  and
      Petukhova, Kseniia  and
      Kochmar, Ekaterina",
    editor = "Chiruzzo, Luis  and
      Ritter, Alan  and
      Wang, Lu",
    booktitle = "Proceedings of the 2025 Conference of the Nations of the Americas Chapter of the Association for Computational Linguistics: Human Language Technologies (Volume 1: Long Papers)",
    month = apr,
    year = "2025",
    address = "Albuquerque, New Mexico",
    publisher = "Association for Computational Linguistics",
    url = "https://aclanthology.org/2025.naacl-long.57/",
    doi = "10.18653/v1/2025.naacl-long.57",
    pages = "1234--1251",
    ISBN = "979-8-89176-189-6"
}

@inproceedings{education-q-2025,
    title = "{E}ducation{Q}: Evaluating {LLM}s' Teaching Capabilities Through Multi-Agent Dialogue Framework",
    author = "Shi, Yao  and
      Liang, Rongkeng  and
      Xu, Yong",
    editor = "Che, Wanxiang  and
      Nabende, Joyce  and
      Shutova, Ekaterina  and
      Pilehvar, Mohammad Taher",
    booktitle = "Proceedings of the 63rd Annual Meeting of the Association for Computational Linguistics (Volume 1: Long Papers)",
    month = jul,
    year = "2025",
    address = "Vienna, Austria",
    publisher = "Association for Computational Linguistics",
    url = "https://aclanthology.org/2025.acl-long.1576/",
    doi = "10.18653/v1/2025.acl-long.1576",
    pages = "32799--32828",
    ISBN = "979-8-89176-251-0"
}

@article{math-tutor-2025,
  title={MathTutorBench: A Benchmark for Measuring Open-ended Pedagogical Capabilities of LLM Tutors},
  author={Jakub Macina and Nico Daheim and Ido Hakimi and Manu Kapur and Iryna Gurevych and Mrinmaya Sachan},
  journal={ArXiv},
  year={2025},
  volume={abs/2502.18940},
  url={https://api.semanticscholar.org/CorpusID:276617751}
}

@inproceedings{traver-2025,
    title = "Training Turn-by-Turn Verifiers for Dialogue Tutoring Agents: The Curious Case of {LLM}s as Your Coding Tutors",
    author = "Wang, Jian  and
      Dai, Yinpei  and
      Zhang, Yichi  and
      Ma, Ziqiao  and
      Li, Wenjie  and
      Chai, Joyce",
    editor = "Che, Wanxiang  and
      Nabende, Joyce  and
      Shutova, Ekaterina  and
      Pilehvar, Mohammad Taher",
    booktitle = "Findings of the Association for Computational Linguistics: ACL 2025",
    month = jul,
    year = "2025",
    address = "Vienna, Austria",
    publisher = "Association for Computational Linguistics",
    url = "https://aclanthology.org/2025.findings-acl.642/",
    doi = "10.18653/v1/2025.findings-acl.642",
    pages = "12416--12436",
    ISBN = "979-8-89176-256-5"
}

@inproceedings{eli-why-2025,
    title = "{ELI}-Why: Evaluating the Pedagogical Utility of Language Model Explanations",
    author = "Joshi, Brihi  and
      He, Keyu  and
      Ramnath, Sahana  and
      Sabouri, Sadra  and
      Zhou, Kaitlyn  and
      Chattopadhyay, Souti  and
      Swayamdipta, Swabha  and
      Ren, Xiang",
    editor = "Che, Wanxiang  and
      Nabende, Joyce  and
      Shutova, Ekaterina  and
      Pilehvar, Mohammad Taher",
    booktitle = "Findings of the Association for Computational Linguistics: ACL 2025",
    month = jul,
    year = "2025",
    address = "Vienna, Austria",
    publisher = "Association for Computational Linguistics",
    url = "https://aclanthology.org/2025.findings-acl.1306/",
    doi = "10.18653/v1/2025.findings-acl.1306",
    pages = "25466--25499",
    ISBN = "979-8-89176-256-5"
}
%TC:endignore
\end{document}